\documentclass[prl,floatfix,twocolumn,byrevtex,superscriptaddress]{revtex4-1}
\usepackage[english]{babel}
\usepackage{amsmath}
\usepackage{graphicx}
\usepackage{float}
\usepackage{bm}
\usepackage{multirow}
\usepackage{dcolumn}
\usepackage{hyperref}
\usepackage[normalem]{ulem}
\usepackage[dvipsnames,usenames]{color}

\newcommand{\SIROx}{Sr$_2$Ir$_{1-x}$Rh$_{x}$O$_{4}$}
\newcommand{\SIO}{Sr$_2$IrO$_{4}$}

\begin{document}

\title{Optical Signature of a Crossover from Mott- to Slater-type Gap in Sr$_2$Ir$_{1-x}$Rh$_{x}$O$_{4}$}
\author{B. Xu}
\email[]{bing.xu@unifr.ch}
\affiliation{University of Fribourg, Department of Physics and Fribourg Center for Nanomaterials, Chemin du Mus\'{e}e 3, CH-1700 Fribourg, Switzerland}

\author{P. Marsik}
\author{E. Sheveleva}
\author{F. Lyzwa}
\affiliation{University of Fribourg, Department of Physics and Fribourg Center for Nanomaterials, Chemin du Mus\'{e}e 3, CH-1700 Fribourg, Switzerland}

\author{A. Louat}
\author{V. Brouet}
\affiliation{Laboratoire de Physique des Solides, CNRS, Univ. Paris-Sud, Universit\'{e} Paris-Saclay, 91405 Orsay Cedex, France
}
\author{D. Munzar}
\affiliation{Department of Condensed Matter Physics, Faculty of Science, and Central European Institute of Technology, Masaryk University, Kotl\'{a}\v{r}sk\'{a} 2, 61137 Brno, Czech Republic}

\author{C. Bernhard}
\email[]{christian.bernhard@unifr.ch}
\affiliation{University of Fribourg, Department of Physics and Fribourg Center for Nanomaterials, Chemin du Mus\'{e}e 3, CH-1700 Fribourg, Switzerland}

\date{\today}

%
%

\begin{abstract}
With optical spectroscopy we provide evidence that the insulator-metal transition in Sr$_2$Ir$_{1-x}$Rh$_{x}$O$_{4}$ occurs close to a crossover from the Mott- to the Slater-type. The Mott-gap at $x = 0$ persists to high temperature and evolves without an anomaly across the N\'{e}el temperature, $T_N$. Upon Rh-doping, it collapses rather rapidly and vanishes around $x = 0.055$. Notably, just as the Mott-gap vanishes yet another gap appears that is of the Slater-type and develops right below $T_N$. This Slater-gap is only partial and is accompanied by a reduced scattering rate of the remaining free carriers, similar as in the parent compounds of the iron arsenide superconductors.
\end{abstract}



\maketitle

%
%
The insulator-metal transition (IMT) in complex transition metal oxides with strongly correlated electrons near half-filling of the conduction band is a longstanding research topic~\cite{Imada1998}. In a so-called Mott-insulator the on-site Coulomb-repulsion (U) splits the conduction band into lower and upper Hubbard bands and thus gives rise to a gap in the electronic excitations. A prominent example are the parent compounds of the cuprates, which become high-$T_c$ superconductors upon doping~\cite{Lee2006}. More recently, the layered iridate \SIO\ has obtained great attention as a high-$T_c$ candidate but also as an unusual Mott-insulator~\cite{Wang2011,Kim2012b,Zhang2013,Martins2011,Kim2008,Kim2009,Jackeli2009,Jin2009,Watanabe2010,Arita2012}. It has a relatively weak Coulomb-interaction, due to the extended Ir $5d$ orbitals, but a strong spin-orbit coupling (SOC) that cooperates to induce a Mott-gap. Like the cuprates, it hosts an antiferromagnetic (AF) order which doubles the unit cell and thus can also lead to a splitting of the conduction band. The latter effect dominates for a so-called Slater-insulator for which the gap occurs only below the N\'{e}el temperature $T_N$. Such a Slater-type gap has not yet been observed in these iridates, but the origin of the $J_{\mathrm{eff}} = 1/2$ insulating state is debated and it may be close to a crossover between the Mott- and Slater-limits~\cite{Kim2008,Kim2009,Jackeli2009,Jin2009,Watanabe2010,Arita2012,Martins2011,Ishii2011,Fujiyama2012,Kini2006,Li2013,Yamasaki2014,Singh2018,Hsieh2012,Watanabe2014}.

A transition from a Mott- to a Slater-type insulator thus may be realized when \SIO\ is doped towards a metallic state and the Mott-gap is more rapidly suppressed than the AF order. A promising candidate is \SIROx\ for which the decrease of the dc resistivity upon substitution of Rh for Ir is considerably faster than the one of $T_N$~\cite{Clancy2014,Cao2016,Qi2012}. Several studies have revealed the interesting properties of the IMT in this system~\cite{Cao2016,Lee2012,Cetin2012,Qi2012,Wang2013,Yan2015,Lee2012,Seo2017,Cao2018,Zhou2017,Wang2018,Louat2018,DelaTorre2015,Zwartsenberg2019} which in general can be controlled by a change of (i) the band filling, (ii) the strength of the electronic correlations or (iii) the SOC. Whereas Rh should be isovalent to Ir, it has been shown that the smaller SOC of Rh, as compared to Ir, gives rise to a charge transfer between neighbouring Ir and Rh ions, and thus an effective hole doping~\cite{Cao2016}. A corresponding shift of the chemical potential in \SIROx\ was observed by angle-resolved-photoemission spectroscopy (ARPES)~\cite{Cao2016,Louat2018} which suggest that the IMT occurs around $x = 0.04$~\cite{Cao2016}. On the other hand, magnetization and neutron diffraction measurements establish that the AF transition temperature decreases more gradually and persists up to $x \sim 0.15$~\cite{Cao2016,Ye2013,Ye2015,Qi2012,Brouet2015,Zhao2016,Jeong2017}.

Here we report the optical conductivity across the IMT of \SIROx. In particular, we detail how the Mott-gap of the parent compound evolves with Rh-substitution and gives way to in-gap states and, eventually, a strongly correlated itinerant state. Notably, after the collapse of the Mott-gap we still observe a smaller gap that develops only in the AF state below $T_N$ and thus can be associated with a Slater-type gap. These findings confirm that the IMT of \SIROx\ is governed by a complex interplay of Coulomb repulsion, spin-orbit coupling and antiferromagnetic correlations.

%
%
\begin{figure}[tb]
\includegraphics[width=0.95\columnwidth]{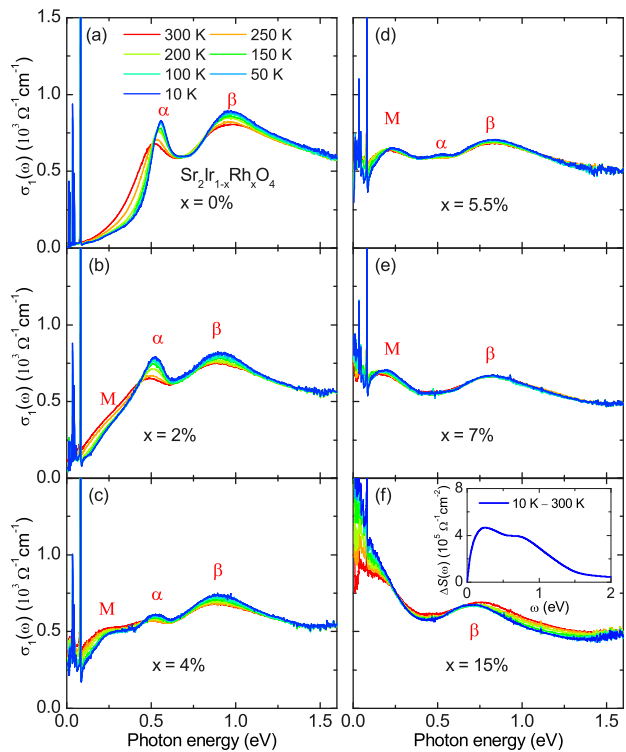}
\caption{ (color online) (a--f) Temperature dependent optical conductivity of \SIROx\ with $0 \leq x \leq 0.15$. Inset of panel (f) shows the difference of the spectral weight integral between 10 and 300~K, $\Delta S(\omega) = S(\omega,\mathrm{10~K}) - S(\omega,\mathrm{300~K})$.}
\label{Fig1}
\end{figure}
Sample synthesis and characterization, as well as experimental methods, are described in Ref.~\footnote{See Supplemental Material at \url{http://link.aps.org/supplemental/xxx} for the details about the characterization of the samples, the reflectivity measurements and the Kramers-Kronig analysis, the Drude-Lorentz fits, as well as the details of temperature-dependent spectra, which includes Refs.~\cite{Homes1993,Dressel2002}.}.

%
Figures~\ref{Fig1}(a--e) display the temperature ($T$) dependent spectra of the real part of the optical conductivity $\sigma_1(\omega)$ for the \SIROx\ crystals with $0 \leq x \leq 0.15$ (corresponding reflectivity spectra can be found in Ref.~\footnotemark[1]). They span the transition from the Mott-insulator at $x = 0$ to a metallic state at $x = 0.15$. The spectra of the parent compound at $x = 0$ in Fig.~\ref{Fig1}(a) reveal an optical gap that develops already above room temperature and deepens continuously toward low temperature. At 10~K, it gives rise to a nearly complete suppression of the electronic conductivity below about 0.2~eV. The sharp features below 0.1~eV are due to infrared-active phonon modes. Towards higher energy, the spectra are dominated by a double peak structure with a lower peak around 0.5~eV ($\alpha$-band) that has been assigned to the transition between the lower and upper Hubbard bands (LHB and UHB) of the half-filled band derived from the $J_\mathrm{eff} = 1/2$ states (around the $X$-point of the Brillouin zone) and an upper peak around 1~eV ($\beta$-band) due to transitions from the $J_\mathrm{eff} = 3/2$ band to the UHB of the $J_\mathrm{eff} = 1/2$ manifold near the $\Gamma$-point~\cite{Zhang2013,Moon2009,Propper2016}. Alternatively, the two-peak structure was interpreted in terms of a mixing of optically active excitations between the LHB and UHB of the $J_\mathrm{eff} = 1/2$ band with an optically forbidden spin-orbit exciton~\cite{Kim2012b,Souri2017}. However, as shown below, this latter interpretation is not supported by the doping dependence of the  $\alpha$- and $\beta$-bands. At $x = 0$, the $\alpha$-band exhibits a strong red-shift with increasing temperature. It also becomes broader and gains a substantial amount of spectral weight from the $\beta$-band. These trends were previously noted and discussed in terms of excitonic and electron-phonon coupling effects, in analogy to those of the parent compounds of the cuprates~\cite{Moon2009,Falck1992}. Notably, the Mott-gap feature evolves continuously below 300~K without any clear anomaly around $T_N \approx$ 255~K.

The above described gap features are still evident in the spectra of the weakly Rh-doped samples with $x =$ 0.02 and 0.04, in Figs.~\ref{Fig1}(b) and ~\ref{Fig1}(c), respectively. However, the spectral weight of the in-gap states increases rather rapidly and their depletion at low temperature remains increasingly incomplete. Most of the additional in-gap spectral weight originates from the $\alpha$-band which shows only a moderate red-shift but exhibits a rapid spectral weight loss and has essentially vanished at $x = 0.055$. This characteristic collapse of the $\alpha$-band signifies the disappearance of the Mott-gap of the $J_\mathrm{eff} = 1/2$ band around $x = 0.055$. In clear contrast, the $\beta$-band evolves rather continuously across $x = 0.055$ and persists up to $x = 0.15$. This suggests that it has a different origin than the $\alpha$-band, in agreement with the assignment given in the previous paragraph. Despite the collapse of the Mott-gap around $x = 0.055$, the low-energy Drude-peak representing the response of itinerant charge carriers remains extremely weak up to $x = 0.07$. Instead, most of the low-energy spectral weight is contained in a mid-infrared band that is in the following denoted as $M$-band.

The observed rapid spectral weight loss and collapse of the Mott-gap related $\alpha$-band associated with the corresponding increase of the low-energy spectral weight, with the eventual formation of a Drude-like peak, is the spectroscopic hallmark of a transition from a Mott-insulator to a strongly correlated metal. The trends are well known from experimental studies of the doping dependent optical response of cuprates~\cite{Imada1998,Lee2005,Lupi2009,Uchida1991}. And, importantly, they are reproduced by calculations using the one band Hubbard model~\cite{Stephan1990,Dagotto1992,Dagotto1994,Nakano2007}. The computed spectra even display the $M$-band, that can be qualitatively interpreted using the high $U$ limit of the one Hubbard model, i.e., the $t$-$J$ model~\cite{Stephan1990,Dagotto1994}. Within this model, the doped holes can be viewed as spin polarons. While the Drude-like peak reflects intraband absorption within the polaron band, the $M$-band is due to more complicated excited states involving magnons. The energy of the $M$-band is predicted to scale with the superexchange constant $J$~\cite{Stephan1990}. In good agreement, since $J$ is about 135~meV in the cuprates~\cite{Coldea2001} and about 60~meV for the present iridate~\cite{Kim2012}, the energy of the $M$-band in the strongly underdoped cuprates at 0.5--0.6~eV~\cite{Uchida1991,Lee2005} is about twice as large as that of the iridates at about 0.25~eV. In addition, the $M$-band may be influenced by electron-phonon interaction, as described for the cuprates in Refs.~\cite{Cappelluti2007,Cappelluti2009,Mishchenko2008}. At very low doping the spectral weight of the $M$-band is predicted to be larger than that of the Drude component~\cite{Nakano2007}. Clearly, the data are consistent also with this prediction. The suppression of the Drude peak for \SIROx\ with $x \sim 0.055$ and 0.07 is likely due to a localization of the polarons by structural disorder (e.g., due to the Rh doping), similar as in weakly doped cuprates~\cite{Lupi2009}.

\begin{figure}[tb]
\includegraphics[width=\columnwidth]{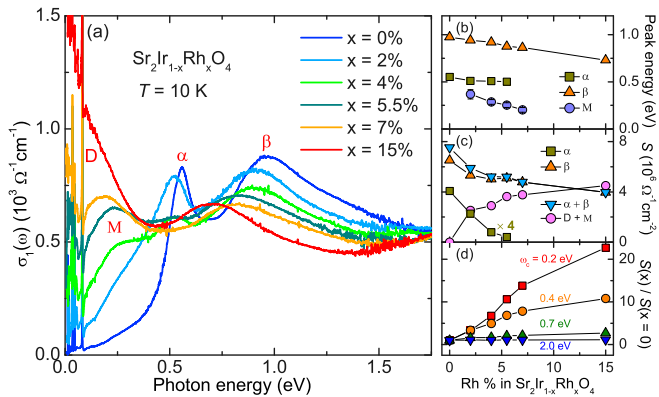}
\caption{ (color online) (a) Comparison of the optical conductivity at 10~K for \SIROx\ at different doping levels. The peak energies and oscillator strengths of the Drude-peak (D) as well as the $\alpha$-, $\beta$- and $M$-bands to the spectra in (a) are shown in (b) and (c), respectively. (d) Doping dependence of the normalized spectral weight for different cutoff frequencies $\omega_c$.}
\label{Fig2}
\end{figure}

The characteristic changes connected with the IMT are further detailed in Fig.~\ref{Fig2}(a), which highlights that $\sigma_1(\omega)$ at 10~K increases rather rapidly with Rh-doping below about 0.4~eV, whereas it decreases at a similar rate above 0.4 eV (up to about 1.8 eV). By fitting the $\sigma_1(\omega)$ spectra with a Drude-Lorentz model (see details in Ref.~\footnotemark[1]), we have quantified the contributions of the free carrier Drude-response ($D$) and the $M$-, $\alpha$-, and $\beta$-bands. The derived peak positions and oscillator strengths are displayed in Figs.~\ref{Fig2}(b) and \ref{Fig2}(c), respectively. The energy of the $\beta$-peak decreases continuously from about 1~eV at $x = 0$ to about 0.75~eV at $x = 0.15$. The $\alpha$-peak shows a similar, moderate redshift, but its spectral weight decreases rapidly and vanishes around $x = 0.055$, marking a sudden collapse of the Mott-gap. On the other hand, the M-peak first appears at $x = 0.02$ and softens and eventually merges with the Drude-peak around $x = 0.15$. The $D$- and $M$-bands together gain about the same amount of spectral weight as the one lost by the $\alpha$- and $\beta$-bands.

The doping dependent spectral weight redistribution has been further analysed in terms of the quantity, $S(\omega_c) = \int^{\omega_c}_0\sigma_{1}(\omega) d\omega$, calculated for different cut-off frequencies, $\omega_c$. Figure~\ref{Fig2}(d) shows the normalized spectral weight $S(\omega_c, x)/S(\omega_c, x = 0)$ for the spectra at 10~K for some representative values of $\omega_c$. For the smallest value of 0.2~eV that is well within the gap region at $x = 0$, the spectral weight increases very rapidly with the Rh content and continues increasing up to $x = 0.15$. For the higher cut-off frequencies, this increase becomes less pronounced until at $\omega_c =$ 2~eV there is no more obvious change. This confirms that the IMT mostly involves electronic excitations below 2~eV.

\begin{figure}[tb]
\includegraphics[width=\columnwidth]{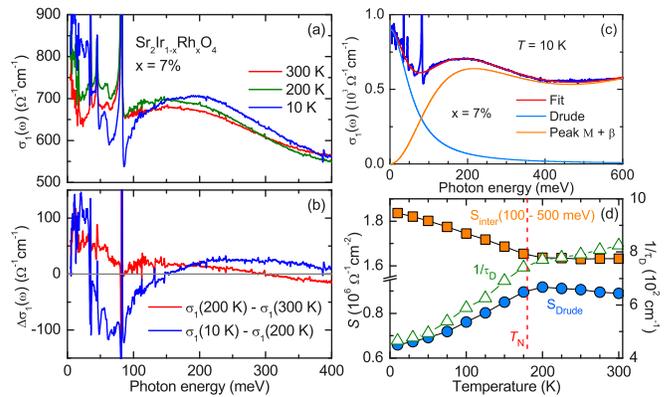}
\caption{ (color online) (a) Optical conductivity of the $x = 0.07$ sample at representative temperatures well above, around  and well below $T_N \approx$ 180~K. (b) The corresponding difference plots of the spectra of panel (a). (c) Drude-Lorentz fit of the $\sigma_1(\omega)$ spectrum at 10~K. (d) Temperature dependences of the Drude weight (circles) and scattering rate (triangles), as well as of the non-Drude spectral weight between 100 and 500~meV (squares).}
\label{Fig3}
\end{figure}

Next, we focus on the gap formation of the $x = 0.07$ sample that occurs right below $T_N$ and thus seems to be of the Slater- rather than the Mott-type (a similar trend occurs for the $x = 0.055$ sample). Figures~\ref{Fig3}(a) and (b) show representative $\sigma_1(\omega)$ spectra well above, around  and well below $T_N \approx$ 180~K and the corresponding difference plots, respectively. The low-energy part of the spectrum at 300~K is composed of a prominent $M$-band with a broad maximum around 150~meV and a weaker and fairly broad Drude-response (that is superimposed by several sharp phonon modes). In the paramagnetic state, between 300 and 200~K, the spectral weight of the $M$-band and the Drude-peak increases by a moderate amount. This additional low-energy spectral weight is accumulated from a broad energy range above 0.4~eV and the overall shape of the spectra is characteristic of the response of a bad metal~\cite{Takenaka2002}. To the contrary, when going from $T_N$ to $T = \mathrm{10~K} << T_N$ a gap-like feature appears that involves a partial suppression of the conductivity in the range from 40 to 160~meV. Most of the missing spectral weight is transferred above the gap edge where the conductivity is enhanced up to about 400~meV. A smaller part is shifted to low energy, i.e., $<$ 40~meV, where it contributes to a narrowed Drude-peak. These spectral changes below $T_N$ are characteristic of the formation of a partial energy gap in the electronic excitations with a magnitude of $2\Delta \simeq$ 160~meV. Similar trends have been reported for other materials with an AF order that involves spins of itinerant charge carriers. A prominent example of such a spin-density-wave (SDW) material are the parent compounds of the iron arsenide superconductors~\cite{Hu2008,Mallett2017PRB,Xu2018,Charnukha2013}.

The analogy with a partial SDW gap state is further substantiated by a fit with a Drude-Lorentz model. Figure~\ref{Fig3}(c) shows for the 10~K spectrum (blue line) that a good fit (red line) is obtained with the sum of a Drude-term (light-blue line) and two Lorentz-oscillators (orange line) that represent the contributions of the $M$- and $\beta$-bands. The temperature evolution of the obtained parameters in Fig.~\ref{Fig3}(d) confirms that the spectral weight and the scattering rate of the Drude-response exhibit clear anomalies at $T_N$. Whereas the Drude weight slightly increases between 300 and 200~K, it suddenly starts to decrease below $T_N$, at 10~K it is reduced by about 30\% as compared to the value at $T_N$. Likewise, the scattering rate of the free carriers decreases more strongly below $T_N$ than above $T_N$. The orange symbols in Fig.~\ref{Fig3}(d) show that a similar amount of spectral weight, as the one lost by the Drude-peak, is gained below $T_N$ by the incoherent part of the spectrum in the energy range between 100 and 500~meV. This confirms that, similar to the iron arsenide parent compounds, the SDW gap below $T_N$ leads to spectral weight transfer from low energy to the region above the gap edge. The enhanced conductivity at very low energies arises from the strong reduction of the scattering rate of the free carriers of the remaining parts of the Fermi-surface.

\begin{figure}[tb]
\includegraphics[width=\columnwidth]{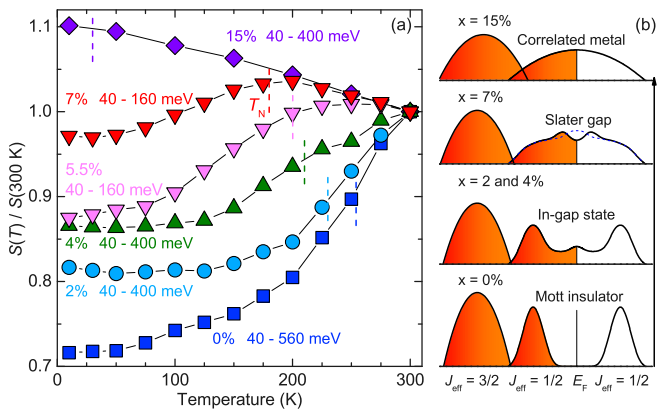}
\caption{ (color online) (a) Temperature dependence of the low-energy spectral weight showing the progression of the gap formations at $0 \leq x \leq 0.07$ and the coherent metallic response at $x = 0.15$. Vertical dashed lines denote $T_N$. (b) Schematic diagram of the electronic states near the Fermi level ($E_\mathrm{F}$) for \SIROx. At $x = 0$, a Mott-gap occurs in the half-filled $J_{\mathrm{eff}} = 1/2$ band whereas the $J_{\mathrm{eff}} = 3/2$ is fully occupied. Rh-doping of $x =$ 0.02 and 0.04 produces in-gap states. At $x = 0.07$, the Mott gap has collapsed but the states near the Fermi level are still partially suppressed by a Slater gap that is induced by the antiferromagnetic order. Finally, at $x = 0.15$, a correlated metal appears.}
\label{Fig4}
\end{figure}
Next, we compare in Fig.~\ref{Fig4}(a) the temperature evolution of the low-energy spectral weight (normalized to the value at 300~K) from $x =$ 0 to 0.15. For $x \leq 0.04 $ it confirms that the gap formation evolves without any clear anomaly across $T_N$. At $x = 0.055$, the spectral weight is almost temperature independent above $T_N \approx$ 200~K, but below $T_N$ it decreases noticeably. This shows that just as the Mott-gap collapses the AF correlations start to contribute to the gapping of the low-energy excitations. Moreover, it suggests that the Mott-type and Slater-type mechanisms are cooperative. The most pronounced anomaly around $T_N$ is seen at $x = 0.07$, where the low-energy spectral weight increases at first between 300 and 200~K before it suddenly starts to decrease below $T_N \approx$ 180~K. The increase of the low-energy spectral weight at $T > T_N$ together with the absence of an $\alpha$-peak confirm the complete suppression of the Mott-gap at $x = 0.07$. The crossover from a Mott-type to a Slater-type gap thus occurs around $x = 0.055$.

At $x = 0.15$ the low-energy spectral weight increases continuously toward low temperature, as expected for a correlated metal. No sign of a gap formation is seen here down to the lowest temperature and frequency measured. Below the lower limit of our measurement of $\sim$ 10~meV, there may still be a pseudogap, as it has been observed in ARPES experiments~\cite{Cao2016,Louat2018}. Indeed, the dc resistivity data~\footnotemark[1] are consistent with a decrease of the conductivity below $\sim$ 10~meV. A weak localization of (some of) the charge carriers can also be the reason why the IMT in the transport data occurs close to $x = 0.07$, rather than at the collapse of the Mott-gap around $x = 0.055$.

Finally, note that the free carrier response in the metallic state at $x = 0.15$ in Figure~\ref{Fig1}(f) is strongly temperature dependent. With decreasing temperature it gains a large amount of spectral weight from the $\beta$-band. The unusually high energy scale of this spectral weight redistribution is evident from the graph of the spectral weight difference, $\Delta S(\omega) = S(\omega,10K) - S(\omega,300K)$, in the the inset of  Figure~\ref{Fig1}(f). It reveals a steep rise of $\Delta S(\omega)$ below about 200~meV, due to the spectral weight gain of the Drude-response, that is followed by a more gradual decrease that continues up to about 1.5~eV, above which the spectral weight is almost compensated. A weaker structure around 0.7~eV arises because the spectral weight loss of the $\beta$-band is accompanied by a rather large red-shift of about 100~meV. Notably, the inverse spectral weight shift is observed for the insulating samples at $x < 0.055$ for which the spectral weight of the $\beta$-band increases as the Mott-gap grows and depletes the in-gap states at low $T$. The direction of the spectral weight redistribution between the $\beta$-band and the low-energy excitations changes around $x = 0.07$. For this particular doping, the spectral weight shift is limited to a much lower energy scale on the order of the Slater-gap. While we do not have a conclusive explanation of this remarkable trend, it indicates that the itinerant charge carriers at $x = 0.15$ remain strongly correlated~\cite{Toschi2005}.

On the basis of the above described results and their similarity to those of the cuprates and the results of Hubbard model based calculations, we have arrived at the schematic representation of the doping dependence of the low energy density of states of \SIROx\ shown in Fig.~\ref{Fig4}(b).

%
%
In summary, we studied the temperature- and doping-dependence of the optical conductivity of \SIROx\ crystals with $0 \leq x \leq 0.15$. In particular, we showed how the Mott-gap vanishes and how the in-gap states and eventually a correlated metallic state emerge with increasing Rh-doping. Moreover, we found that, just as the Mott-gap collapses, a second kind of gap appears that develops only in the AF state and thus is of a Slater-type. Our findings suggest that the IMT in \SIROx\ undergoes a crossover from a Mott- to a Slater-type as controlled by the delicate interplay between electronic correlations, spin-orbit coupling, and antiferromagnetic order. Finally, in the metallic state at $x = 0.15$ the Drude-response is also rather unusual since its spectral weight is strongly temperature dependent with an underlying spectral weight transfer that involves a large energy scale of about 1.5 eV and thus is indicative of the persistence of strong correlation effects.

%
%
We acknowledge valuable discussions with A. Georges. Work at the University of Fribourg was supported by the Schweizerische Nationalfonds (SNF) by Grant No. 200020-172611. A. L. and V. B. acknowledge financial support from the French National Agency ANR-15-CE30-0009-01 SOCRATE.

%
%
%


\begin{thebibliography}{64}%
\makeatletter
\providecommand \@ifxundefined [1]{%
 \@ifx{#1\undefined}
}%
\providecommand \@ifnum [1]{%
 \ifnum #1\expandafter \@firstoftwo
 \else \expandafter \@secondoftwo
 \fi
}%
\providecommand \@ifx [1]{%
 \ifx #1\expandafter \@firstoftwo
 \else \expandafter \@secondoftwo
 \fi
}%
\providecommand \natexlab [1]{#1}%
\providecommand \enquote  [1]{``#1''}%
\providecommand \bibnamefont  [1]{#1}%
\providecommand \bibfnamefont [1]{#1}%
\providecommand \citenamefont [1]{#1}%
\providecommand \href@noop [0]{\@secondoftwo}%
\providecommand \href [0]{\begingroup \@sanitize@url \@href}%
\providecommand \@href[1]{\@@startlink{#1}\@@href}%
\providecommand \@@href[1]{\endgroup#1\@@endlink}%
\providecommand \@sanitize@url [0]{\catcode `\\12\catcode `\$12\catcode
  `\&12\catcode `\#12\catcode `\^12\catcode `\_12\catcode `\%12\relax}%
\providecommand \@@startlink[1]{}%
\providecommand \@@endlink[0]{}%
\providecommand \url  [0]{\begingroup\@sanitize@url \@url }%
\providecommand \@url [1]{\endgroup\@href {#1}{\urlprefix }}%
\providecommand \urlprefix  [0]{URL }%
\providecommand \Eprint [0]{\href }%
\providecommand \doibase [0]{http://dx.doi.org/}%
\providecommand \selectlanguage [0]{\@gobble}%
\providecommand \bibinfo  [0]{\@secondoftwo}%
\providecommand \bibfield  [0]{\@secondoftwo}%
\providecommand \translation [1]{[#1]}%
\providecommand \BibitemOpen [0]{}%
\providecommand \bibitemStop [0]{}%
\providecommand \bibitemNoStop [0]{.\EOS\space}%
\providecommand \EOS [0]{\spacefactor3000\relax}%
\providecommand \BibitemShut  [1]{\csname bibitem#1\endcsname}%
\let\auto@bib@innerbib\@empty
\bibitem [{\citenamefont {Imada}\ \emph {et~al.}(1998)\citenamefont {Imada},
  \citenamefont {Fujimori},\ and\ \citenamefont {Tokura}}]{Imada1998}%
  \BibitemOpen
  \bibfield  {author} {\bibinfo {author} {\bibfnamefont {M.}~\bibnamefont
  {Imada}}, \bibinfo {author} {\bibfnamefont {A.}~\bibnamefont {Fujimori}}, \
  and\ \bibinfo {author} {\bibfnamefont {Y.}~\bibnamefont {Tokura}},\ }\href
  {\doibase 10.1103/RevModPhys.70.1039} {\bibfield  {journal} {\bibinfo
  {journal} {Rev. Mod. Phys.}\ }\textbf {\bibinfo {volume} {70}},\ \bibinfo
  {pages} {1039} (\bibinfo {year} {1998})}\BibitemShut {NoStop}%
\bibitem [{\citenamefont {Lee}\ \emph {et~al.}(2006)\citenamefont {Lee},
  \citenamefont {Nagaosa},\ and\ \citenamefont {Wen}}]{Lee2006}%
  \BibitemOpen
  \bibfield  {author} {\bibinfo {author} {\bibfnamefont {P.~A.}\ \bibnamefont
  {Lee}}, \bibinfo {author} {\bibfnamefont {N.}~\bibnamefont {Nagaosa}}, \ and\
  \bibinfo {author} {\bibfnamefont {X.-G.}\ \bibnamefont {Wen}},\ }\href
  {\doibase 10.1103/RevModPhys.78.17} {\bibfield  {journal} {\bibinfo
  {journal} {Rev. Mod. Phys.}\ }\textbf {\bibinfo {volume} {78}},\ \bibinfo
  {pages} {17} (\bibinfo {year} {2006})}\BibitemShut {NoStop}%
\bibitem [{\citenamefont {Wang}\ and\ \citenamefont
  {Senthil}(2011)}]{Wang2011}%
  \BibitemOpen
  \bibfield  {author} {\bibinfo {author} {\bibfnamefont {F.}~\bibnamefont
  {Wang}}\ and\ \bibinfo {author} {\bibfnamefont {T.}~\bibnamefont {Senthil}},\
  }\href {\doibase 10.1103/PhysRevLett.106.136402} {\bibfield  {journal}
  {\bibinfo  {journal} {Phys. Rev. Lett.}\ }\textbf {\bibinfo {volume} {106}},\
  \bibinfo {pages} {136402} (\bibinfo {year} {2011})}\BibitemShut {NoStop}%
\bibitem [{\citenamefont {Kim}\ \emph {et~al.}(2012{\natexlab{a}})\citenamefont
  {Kim}, \citenamefont {Khaliullin},\ and\ \citenamefont {Min}}]{Kim2012b}%
  \BibitemOpen
  \bibfield  {author} {\bibinfo {author} {\bibfnamefont {B.~H.}\ \bibnamefont
  {Kim}}, \bibinfo {author} {\bibfnamefont {G.}~\bibnamefont {Khaliullin}}, \
  and\ \bibinfo {author} {\bibfnamefont {B.~I.}\ \bibnamefont {Min}},\ }\href
  {\doibase 10.1103/PhysRevLett.109.167205} {\bibfield  {journal} {\bibinfo
  {journal} {Phys. Rev. Lett.}\ }\textbf {\bibinfo {volume} {109}},\ \bibinfo
  {pages} {167205} (\bibinfo {year} {2012}{\natexlab{a}})}\BibitemShut
  {NoStop}%
\bibitem [{\citenamefont {Zhang}\ \emph {et~al.}(2013)\citenamefont {Zhang},
  \citenamefont {Haule},\ and\ \citenamefont {Vanderbilt}}]{Zhang2013}%
  \BibitemOpen
  \bibfield  {author} {\bibinfo {author} {\bibfnamefont {H.}~\bibnamefont
  {Zhang}}, \bibinfo {author} {\bibfnamefont {K.}~\bibnamefont {Haule}}, \ and\
  \bibinfo {author} {\bibfnamefont {D.}~\bibnamefont {Vanderbilt}},\ }\href
  {\doibase 10.1103/PhysRevLett.111.246402} {\bibfield  {journal} {\bibinfo
  {journal} {Phys. Rev. Lett.}\ }\textbf {\bibinfo {volume} {111}},\ \bibinfo
  {pages} {246402} (\bibinfo {year} {2013})}\BibitemShut {NoStop}%
\bibitem [{\citenamefont {Martins}\ \emph {et~al.}(2011)\citenamefont
  {Martins}, \citenamefont {Aichhorn}, \citenamefont {Vaugier},\ and\
  \citenamefont {Biermann}}]{Martins2011}%
  \BibitemOpen
  \bibfield  {author} {\bibinfo {author} {\bibfnamefont {C.}~\bibnamefont
  {Martins}}, \bibinfo {author} {\bibfnamefont {M.}~\bibnamefont {Aichhorn}},
  \bibinfo {author} {\bibfnamefont {L.}~\bibnamefont {Vaugier}}, \ and\
  \bibinfo {author} {\bibfnamefont {S.}~\bibnamefont {Biermann}},\ }\href
  {\doibase 10.1103/PhysRevLett.107.266404} {\bibfield  {journal} {\bibinfo
  {journal} {Phys. Rev. Lett.}\ }\textbf {\bibinfo {volume} {107}},\ \bibinfo
  {pages} {266404} (\bibinfo {year} {2011})}\BibitemShut {NoStop}%
\bibitem [{\citenamefont {Kim}\ \emph {et~al.}(2008)\citenamefont {Kim},
  \citenamefont {Jin}, \citenamefont {Moon}, \citenamefont {Kim}, \citenamefont
  {Park}, \citenamefont {Leem}, \citenamefont {Yu}, \citenamefont {Noh},
  \citenamefont {Kim}, \citenamefont {Oh}, \citenamefont {Park}, \citenamefont
  {Durairaj}, \citenamefont {Cao},\ and\ \citenamefont {Rotenberg}}]{Kim2008}%
  \BibitemOpen
  \bibfield  {author} {\bibinfo {author} {\bibfnamefont {B.~J.}\ \bibnamefont
  {Kim}}, \bibinfo {author} {\bibfnamefont {H.}~\bibnamefont {Jin}}, \bibinfo
  {author} {\bibfnamefont {S.~J.}\ \bibnamefont {Moon}}, \bibinfo {author}
  {\bibfnamefont {J.-Y.}\ \bibnamefont {Kim}}, \bibinfo {author} {\bibfnamefont
  {B.-G.}\ \bibnamefont {Park}}, \bibinfo {author} {\bibfnamefont {C.~S.}\
  \bibnamefont {Leem}}, \bibinfo {author} {\bibfnamefont {J.}~\bibnamefont
  {Yu}}, \bibinfo {author} {\bibfnamefont {T.~W.}\ \bibnamefont {Noh}},
  \bibinfo {author} {\bibfnamefont {C.}~\bibnamefont {Kim}}, \bibinfo {author}
  {\bibfnamefont {S.-J.}\ \bibnamefont {Oh}}, \bibinfo {author} {\bibfnamefont
  {J.-H.}\ \bibnamefont {Park}}, \bibinfo {author} {\bibfnamefont
  {V.}~\bibnamefont {Durairaj}}, \bibinfo {author} {\bibfnamefont
  {G.}~\bibnamefont {Cao}}, \ and\ \bibinfo {author} {\bibfnamefont
  {E.}~\bibnamefont {Rotenberg}},\ }\href {\doibase
  10.1103/PhysRevLett.101.076402} {\bibfield  {journal} {\bibinfo  {journal}
  {Phys. Rev. Lett.}\ }\textbf {\bibinfo {volume} {101}},\ \bibinfo {pages}
  {076402} (\bibinfo {year} {2008})}\BibitemShut {NoStop}%
\bibitem [{\citenamefont {Kim}\ \emph {et~al.}(2009)\citenamefont {Kim},
  \citenamefont {Ohsumi}, \citenamefont {Komesu}, \citenamefont {Sakai},
  \citenamefont {Morita}, \citenamefont {Takagi},\ and\ \citenamefont
  {Arima}}]{Kim2009}%
  \BibitemOpen
  \bibfield  {author} {\bibinfo {author} {\bibfnamefont {B.~J.}\ \bibnamefont
  {Kim}}, \bibinfo {author} {\bibfnamefont {H.}~\bibnamefont {Ohsumi}},
  \bibinfo {author} {\bibfnamefont {T.}~\bibnamefont {Komesu}}, \bibinfo
  {author} {\bibfnamefont {S.}~\bibnamefont {Sakai}}, \bibinfo {author}
  {\bibfnamefont {T.}~\bibnamefont {Morita}}, \bibinfo {author} {\bibfnamefont
  {H.}~\bibnamefont {Takagi}}, \ and\ \bibinfo {author} {\bibfnamefont
  {T.}~\bibnamefont {Arima}},\ }\href {\doibase 10.1126/science.1167106}
  {\bibfield  {journal} {\bibinfo  {journal} {Science}\ }\textbf {\bibinfo
  {volume} {323}},\ \bibinfo {pages} {1329} (\bibinfo {year}
  {2009})}\BibitemShut {NoStop}%
\bibitem [{\citenamefont {Jackeli}\ and\ \citenamefont
  {Khaliullin}(2009)}]{Jackeli2009}%
  \BibitemOpen
  \bibfield  {author} {\bibinfo {author} {\bibfnamefont {G.}~\bibnamefont
  {Jackeli}}\ and\ \bibinfo {author} {\bibfnamefont {G.}~\bibnamefont
  {Khaliullin}},\ }\href {\doibase 10.1103/PhysRevLett.102.017205} {\bibfield
  {journal} {\bibinfo  {journal} {Phys. Rev. Lett.}\ }\textbf {\bibinfo
  {volume} {102}},\ \bibinfo {pages} {017205} (\bibinfo {year}
  {2009})}\BibitemShut {NoStop}%
\bibitem [{\citenamefont {Jin}\ \emph {et~al.}(2009)\citenamefont {Jin},
  \citenamefont {Jeong}, \citenamefont {Ozaki},\ and\ \citenamefont
  {Yu}}]{Jin2009}%
  \BibitemOpen
  \bibfield  {author} {\bibinfo {author} {\bibfnamefont {H.}~\bibnamefont
  {Jin}}, \bibinfo {author} {\bibfnamefont {H.}~\bibnamefont {Jeong}}, \bibinfo
  {author} {\bibfnamefont {T.}~\bibnamefont {Ozaki}}, \ and\ \bibinfo {author}
  {\bibfnamefont {J.}~\bibnamefont {Yu}},\ }\href {\doibase
  10.1103/PhysRevB.80.075112} {\bibfield  {journal} {\bibinfo  {journal} {Phys.
  Rev. B}\ }\textbf {\bibinfo {volume} {80}},\ \bibinfo {pages} {075112}
  (\bibinfo {year} {2009})}\BibitemShut {NoStop}%
\bibitem [{\citenamefont {Watanabe}\ \emph {et~al.}(2010)\citenamefont
  {Watanabe}, \citenamefont {Shirakawa},\ and\ \citenamefont
  {Yunoki}}]{Watanabe2010}%
  \BibitemOpen
  \bibfield  {author} {\bibinfo {author} {\bibfnamefont {H.}~\bibnamefont
  {Watanabe}}, \bibinfo {author} {\bibfnamefont {T.}~\bibnamefont {Shirakawa}},
  \ and\ \bibinfo {author} {\bibfnamefont {S.}~\bibnamefont {Yunoki}},\ }\href
  {\doibase 10.1103/PhysRevLett.105.216410} {\bibfield  {journal} {\bibinfo
  {journal} {Phys. Rev. Lett.}\ }\textbf {\bibinfo {volume} {105}},\ \bibinfo
  {pages} {216410} (\bibinfo {year} {2010})}\BibitemShut {NoStop}%
\bibitem [{\citenamefont {Arita}\ \emph {et~al.}(2012)\citenamefont {Arita},
  \citenamefont {Kune\ifmmode~\check{s}\else \v{s}\fi{}}, \citenamefont
  {Kozhevnikov}, \citenamefont {Eguiluz},\ and\ \citenamefont
  {Imada}}]{Arita2012}%
  \BibitemOpen
  \bibfield  {author} {\bibinfo {author} {\bibfnamefont {R.}~\bibnamefont
  {Arita}}, \bibinfo {author} {\bibfnamefont {J.}~\bibnamefont
  {Kune\ifmmode~\check{s}\else \v{s}\fi{}}}, \bibinfo {author} {\bibfnamefont
  {A.~V.}\ \bibnamefont {Kozhevnikov}}, \bibinfo {author} {\bibfnamefont
  {A.~G.}\ \bibnamefont {Eguiluz}}, \ and\ \bibinfo {author} {\bibfnamefont
  {M.}~\bibnamefont {Imada}},\ }\href {\doibase 10.1103/PhysRevLett.108.086403}
  {\bibfield  {journal} {\bibinfo  {journal} {Phys. Rev. Lett.}\ }\textbf
  {\bibinfo {volume} {108}},\ \bibinfo {pages} {086403} (\bibinfo {year}
  {2012})}\BibitemShut {NoStop}%
\bibitem [{\citenamefont {Ishii}\ \emph {et~al.}(2011)\citenamefont {Ishii},
  \citenamefont {Jarrige}, \citenamefont {Yoshida}, \citenamefont {Ikeuchi},
  \citenamefont {Mizuki}, \citenamefont {Ohashi}, \citenamefont {Takayama},
  \citenamefont {Matsuno},\ and\ \citenamefont {Takagi}}]{Ishii2011}%
  \BibitemOpen
  \bibfield  {author} {\bibinfo {author} {\bibfnamefont {K.}~\bibnamefont
  {Ishii}}, \bibinfo {author} {\bibfnamefont {I.}~\bibnamefont {Jarrige}},
  \bibinfo {author} {\bibfnamefont {M.}~\bibnamefont {Yoshida}}, \bibinfo
  {author} {\bibfnamefont {K.}~\bibnamefont {Ikeuchi}}, \bibinfo {author}
  {\bibfnamefont {J.}~\bibnamefont {Mizuki}}, \bibinfo {author} {\bibfnamefont
  {K.}~\bibnamefont {Ohashi}}, \bibinfo {author} {\bibfnamefont
  {T.}~\bibnamefont {Takayama}}, \bibinfo {author} {\bibfnamefont
  {J.}~\bibnamefont {Matsuno}}, \ and\ \bibinfo {author} {\bibfnamefont
  {H.}~\bibnamefont {Takagi}},\ }\href {\doibase 10.1103/PhysRevB.83.115121}
  {\bibfield  {journal} {\bibinfo  {journal} {Phys. Rev. B}\ }\textbf {\bibinfo
  {volume} {83}},\ \bibinfo {pages} {115121} (\bibinfo {year}
  {2011})}\BibitemShut {NoStop}%
\bibitem [{\citenamefont {Fujiyama}\ \emph {et~al.}(2012)\citenamefont
  {Fujiyama}, \citenamefont {Ohsumi}, \citenamefont {Komesu}, \citenamefont
  {Matsuno}, \citenamefont {Kim}, \citenamefont {Takata}, \citenamefont
  {Arima},\ and\ \citenamefont {Takagi}}]{Fujiyama2012}%
  \BibitemOpen
  \bibfield  {author} {\bibinfo {author} {\bibfnamefont {S.}~\bibnamefont
  {Fujiyama}}, \bibinfo {author} {\bibfnamefont {H.}~\bibnamefont {Ohsumi}},
  \bibinfo {author} {\bibfnamefont {T.}~\bibnamefont {Komesu}}, \bibinfo
  {author} {\bibfnamefont {J.}~\bibnamefont {Matsuno}}, \bibinfo {author}
  {\bibfnamefont {B.~J.}\ \bibnamefont {Kim}}, \bibinfo {author} {\bibfnamefont
  {M.}~\bibnamefont {Takata}}, \bibinfo {author} {\bibfnamefont
  {T.}~\bibnamefont {Arima}}, \ and\ \bibinfo {author} {\bibfnamefont
  {H.}~\bibnamefont {Takagi}},\ }\href {\doibase
  10.1103/PhysRevLett.108.247212} {\bibfield  {journal} {\bibinfo  {journal}
  {Phys. Rev. Lett.}\ }\textbf {\bibinfo {volume} {108}},\ \bibinfo {pages}
  {247212} (\bibinfo {year} {2012})}\BibitemShut {NoStop}%
\bibitem [{\citenamefont {Kini}\ \emph {et~al.}(2006)\citenamefont {Kini},
  \citenamefont {Strydom}, \citenamefont {Jeevan}, \citenamefont {Geibel},\
  and\ \citenamefont {Ramakrishnan}}]{Kini2006}%
  \BibitemOpen
  \bibfield  {author} {\bibinfo {author} {\bibfnamefont {N.~S.}\ \bibnamefont
  {Kini}}, \bibinfo {author} {\bibfnamefont {A.~M.}\ \bibnamefont {Strydom}},
  \bibinfo {author} {\bibfnamefont {H.~S.}\ \bibnamefont {Jeevan}}, \bibinfo
  {author} {\bibfnamefont {C.}~\bibnamefont {Geibel}}, \ and\ \bibinfo {author}
  {\bibfnamefont {S.}~\bibnamefont {Ramakrishnan}},\ }\href {\doibase
  10.1088/0953-8984/18/35/008} {\bibfield  {journal} {\bibinfo  {journal}
  {Journal of Physics: Condensed Matter}\ }\textbf {\bibinfo {volume} {18}},\
  \bibinfo {pages} {8205} (\bibinfo {year} {2006})}\BibitemShut {NoStop}%
\bibitem [{\citenamefont {Li}\ \emph {et~al.}(2013)\citenamefont {Li},
  \citenamefont {Cao}, \citenamefont {Okamoto}, \citenamefont {Yi},
  \citenamefont {Lin}, \citenamefont {Sales}, \citenamefont {Yan},
  \citenamefont {Arita}, \citenamefont {Kune{\v{s}}}, \citenamefont
  {Kozhevnikov}, \citenamefont {Eguiluz}, \citenamefont {Imada}, \citenamefont
  {Gai}, \citenamefont {Pan},\ and\ \citenamefont {Mandrus}}]{Li2013}%
  \BibitemOpen
  \bibfield  {author} {\bibinfo {author} {\bibfnamefont {Q.}~\bibnamefont
  {Li}}, \bibinfo {author} {\bibfnamefont {G.}~\bibnamefont {Cao}}, \bibinfo
  {author} {\bibfnamefont {S.}~\bibnamefont {Okamoto}}, \bibinfo {author}
  {\bibfnamefont {J.}~\bibnamefont {Yi}}, \bibinfo {author} {\bibfnamefont
  {W.}~\bibnamefont {Lin}}, \bibinfo {author} {\bibfnamefont {B.~C.}\
  \bibnamefont {Sales}}, \bibinfo {author} {\bibfnamefont {J.}~\bibnamefont
  {Yan}}, \bibinfo {author} {\bibfnamefont {R.}~\bibnamefont {Arita}}, \bibinfo
  {author} {\bibfnamefont {J.}~\bibnamefont {Kune{\v{s}}}}, \bibinfo {author}
  {\bibfnamefont {A.~V.}\ \bibnamefont {Kozhevnikov}}, \bibinfo {author}
  {\bibfnamefont {A.~G.}\ \bibnamefont {Eguiluz}}, \bibinfo {author}
  {\bibfnamefont {M.}~\bibnamefont {Imada}}, \bibinfo {author} {\bibfnamefont
  {Z.}~\bibnamefont {Gai}}, \bibinfo {author} {\bibfnamefont {M.}~\bibnamefont
  {Pan}}, \ and\ \bibinfo {author} {\bibfnamefont {D.~G.}\ \bibnamefont
  {Mandrus}},\ }\href {\doibase 10.1038/srep03073} {\bibfield  {journal}
  {\bibinfo  {journal} {Sci. Rep.}\ }\textbf {\bibinfo {volume} {3}},\ \bibinfo
  {pages} {3073} (\bibinfo {year} {2013})}\BibitemShut {NoStop}%
\bibitem [{\citenamefont {Yamasaki}\ \emph {et~al.}(2014)\citenamefont
  {Yamasaki}, \citenamefont {Tachibana}, \citenamefont {Fujiwara},
  \citenamefont {Higashiya}, \citenamefont {Irizawa}, \citenamefont {Kirilmaz},
  \citenamefont {Pfaff}, \citenamefont {Scheiderer}, \citenamefont {Gabel},
  \citenamefont {Sing}, \citenamefont {Muro}, \citenamefont {Yabashi},
  \citenamefont {Tamasaku}, \citenamefont {Sato}, \citenamefont {Namatame},
  \citenamefont {Taniguchi}, \citenamefont {Hloskovskyy}, \citenamefont
  {Yoshida}, \citenamefont {Okabe}, \citenamefont {Isobe}, \citenamefont
  {Akimitsu}, \citenamefont {Drube}, \citenamefont {Claessen}, \citenamefont
  {Ishikawa}, \citenamefont {Imada}, \citenamefont {Sekiyama},\ and\
  \citenamefont {Suga}}]{Yamasaki2014}%
  \BibitemOpen
  \bibfield  {author} {\bibinfo {author} {\bibfnamefont {A.}~\bibnamefont
  {Yamasaki}}, \bibinfo {author} {\bibfnamefont {S.}~\bibnamefont {Tachibana}},
  \bibinfo {author} {\bibfnamefont {H.}~\bibnamefont {Fujiwara}}, \bibinfo
  {author} {\bibfnamefont {A.}~\bibnamefont {Higashiya}}, \bibinfo {author}
  {\bibfnamefont {A.}~\bibnamefont {Irizawa}}, \bibinfo {author} {\bibfnamefont
  {O.}~\bibnamefont {Kirilmaz}}, \bibinfo {author} {\bibfnamefont
  {F.}~\bibnamefont {Pfaff}}, \bibinfo {author} {\bibfnamefont
  {P.}~\bibnamefont {Scheiderer}}, \bibinfo {author} {\bibfnamefont
  {J.}~\bibnamefont {Gabel}}, \bibinfo {author} {\bibfnamefont
  {M.}~\bibnamefont {Sing}}, \bibinfo {author} {\bibfnamefont {T.}~\bibnamefont
  {Muro}}, \bibinfo {author} {\bibfnamefont {M.}~\bibnamefont {Yabashi}},
  \bibinfo {author} {\bibfnamefont {K.}~\bibnamefont {Tamasaku}}, \bibinfo
  {author} {\bibfnamefont {H.}~\bibnamefont {Sato}}, \bibinfo {author}
  {\bibfnamefont {H.}~\bibnamefont {Namatame}}, \bibinfo {author}
  {\bibfnamefont {M.}~\bibnamefont {Taniguchi}}, \bibinfo {author}
  {\bibfnamefont {A.}~\bibnamefont {Hloskovskyy}}, \bibinfo {author}
  {\bibfnamefont {H.}~\bibnamefont {Yoshida}}, \bibinfo {author} {\bibfnamefont
  {H.}~\bibnamefont {Okabe}}, \bibinfo {author} {\bibfnamefont
  {M.}~\bibnamefont {Isobe}}, \bibinfo {author} {\bibfnamefont
  {J.}~\bibnamefont {Akimitsu}}, \bibinfo {author} {\bibfnamefont
  {W.}~\bibnamefont {Drube}}, \bibinfo {author} {\bibfnamefont
  {R.}~\bibnamefont {Claessen}}, \bibinfo {author} {\bibfnamefont
  {T.}~\bibnamefont {Ishikawa}}, \bibinfo {author} {\bibfnamefont
  {S.}~\bibnamefont {Imada}}, \bibinfo {author} {\bibfnamefont
  {A.}~\bibnamefont {Sekiyama}}, \ and\ \bibinfo {author} {\bibfnamefont
  {S.}~\bibnamefont {Suga}},\ }\href {\doibase 10.1103/PhysRevB.89.121111}
  {\bibfield  {journal} {\bibinfo  {journal} {Phys. Rev. B}\ }\textbf {\bibinfo
  {volume} {89}},\ \bibinfo {pages} {121111} (\bibinfo {year}
  {2014})}\BibitemShut {NoStop}%
\bibitem [{\citenamefont {Singh}\ and\ \citenamefont
  {Pulikkotil}()}]{Singh2018}%
  \BibitemOpen
  \bibfield  {author} {\bibinfo {author} {\bibfnamefont {V.}~\bibnamefont
  {Singh}}\ and\ \bibinfo {author} {\bibfnamefont {J.~J.}\ \bibnamefont
  {Pulikkotil}},\ }\href@noop {} {\enquote {\bibinfo {title} {{Evidence of
  Slater-type mechanism as origin of insulating state in Sr$_2$IrO$_4$}},}\
  }\Eprint {http://arxiv.org/abs/arXiv:1812.06241 (2018)} {arXiv:1812.06241
  (2018)} \BibitemShut {NoStop}%
\bibitem [{\citenamefont {Hsieh}\ \emph {et~al.}(2012)\citenamefont {Hsieh},
  \citenamefont {Mahmood}, \citenamefont {Torchinsky}, \citenamefont {Cao},\
  and\ \citenamefont {Gedik}}]{Hsieh2012}%
  \BibitemOpen
  \bibfield  {author} {\bibinfo {author} {\bibfnamefont {D.}~\bibnamefont
  {Hsieh}}, \bibinfo {author} {\bibfnamefont {F.}~\bibnamefont {Mahmood}},
  \bibinfo {author} {\bibfnamefont {D.~H.}\ \bibnamefont {Torchinsky}},
  \bibinfo {author} {\bibfnamefont {G.}~\bibnamefont {Cao}}, \ and\ \bibinfo
  {author} {\bibfnamefont {N.}~\bibnamefont {Gedik}},\ }\href {\doibase
  10.1103/PhysRevB.86.035128} {\bibfield  {journal} {\bibinfo  {journal} {Phys.
  Rev. B}\ }\textbf {\bibinfo {volume} {86}},\ \bibinfo {pages} {035128}
  (\bibinfo {year} {2012})}\BibitemShut {NoStop}%
\bibitem [{\citenamefont {Watanabe}\ \emph {et~al.}(2014)\citenamefont
  {Watanabe}, \citenamefont {Shirakawa},\ and\ \citenamefont
  {Yunoki}}]{Watanabe2014}%
  \BibitemOpen
  \bibfield  {author} {\bibinfo {author} {\bibfnamefont {H.}~\bibnamefont
  {Watanabe}}, \bibinfo {author} {\bibfnamefont {T.}~\bibnamefont {Shirakawa}},
  \ and\ \bibinfo {author} {\bibfnamefont {S.}~\bibnamefont {Yunoki}},\ }\href
  {\doibase 10.1103/PhysRevB.89.165115} {\bibfield  {journal} {\bibinfo
  {journal} {Phys. Rev. B}\ }\textbf {\bibinfo {volume} {89}},\ \bibinfo
  {pages} {165115} (\bibinfo {year} {2014})}\BibitemShut {NoStop}%
\bibitem [{\citenamefont {Clancy}\ \emph {et~al.}(2014)\citenamefont {Clancy},
  \citenamefont {Lupascu}, \citenamefont {Gretarsson}, \citenamefont {Islam},
  \citenamefont {Hu}, \citenamefont {Casa}, \citenamefont {Nelson},
  \citenamefont {LaMarra}, \citenamefont {Cao},\ and\ \citenamefont
  {Kim}}]{Clancy2014}%
  \BibitemOpen
  \bibfield  {author} {\bibinfo {author} {\bibfnamefont {J.~P.}\ \bibnamefont
  {Clancy}}, \bibinfo {author} {\bibfnamefont {A.}~\bibnamefont {Lupascu}},
  \bibinfo {author} {\bibfnamefont {H.}~\bibnamefont {Gretarsson}}, \bibinfo
  {author} {\bibfnamefont {Z.}~\bibnamefont {Islam}}, \bibinfo {author}
  {\bibfnamefont {Y.~F.}\ \bibnamefont {Hu}}, \bibinfo {author} {\bibfnamefont
  {D.}~\bibnamefont {Casa}}, \bibinfo {author} {\bibfnamefont {C.~S.}\
  \bibnamefont {Nelson}}, \bibinfo {author} {\bibfnamefont {S.~C.}\
  \bibnamefont {LaMarra}}, \bibinfo {author} {\bibfnamefont {G.}~\bibnamefont
  {Cao}}, \ and\ \bibinfo {author} {\bibfnamefont {Y.-J.}\ \bibnamefont
  {Kim}},\ }\href {\doibase 10.1103/PhysRevB.89.054409} {\bibfield  {journal}
  {\bibinfo  {journal} {Phys. Rev. B}\ }\textbf {\bibinfo {volume} {89}},\
  \bibinfo {pages} {054409} (\bibinfo {year} {2014})}\BibitemShut {NoStop}%
\bibitem [{\citenamefont {Cao}\ \emph {et~al.}(2016)\citenamefont {Cao},
  \citenamefont {Wang}, \citenamefont {Waugh}, \citenamefont {Reber},
  \citenamefont {Li}, \citenamefont {Zhou}, \citenamefont {Parham},
  \citenamefont {Park}, \citenamefont {Plumb}, \citenamefont {Rotenberg},
  \citenamefont {Bostwick}, \citenamefont {Denlinger}, \citenamefont {Qi},
  \citenamefont {Hermele}, \citenamefont {Cao},\ and\ \citenamefont
  {Dessau}}]{Cao2016}%
  \BibitemOpen
  \bibfield  {author} {\bibinfo {author} {\bibfnamefont {Y.}~\bibnamefont
  {Cao}}, \bibinfo {author} {\bibfnamefont {Q.}~\bibnamefont {Wang}}, \bibinfo
  {author} {\bibfnamefont {J.~A.}\ \bibnamefont {Waugh}}, \bibinfo {author}
  {\bibfnamefont {T.~J.}\ \bibnamefont {Reber}}, \bibinfo {author}
  {\bibfnamefont {H.}~\bibnamefont {Li}}, \bibinfo {author} {\bibfnamefont
  {X.}~\bibnamefont {Zhou}}, \bibinfo {author} {\bibfnamefont {S.}~\bibnamefont
  {Parham}}, \bibinfo {author} {\bibfnamefont {S.-R.}\ \bibnamefont {Park}},
  \bibinfo {author} {\bibfnamefont {N.~C.}\ \bibnamefont {Plumb}}, \bibinfo
  {author} {\bibfnamefont {E.}~\bibnamefont {Rotenberg}}, \bibinfo {author}
  {\bibfnamefont {A.}~\bibnamefont {Bostwick}}, \bibinfo {author}
  {\bibfnamefont {J.~D.}\ \bibnamefont {Denlinger}}, \bibinfo {author}
  {\bibfnamefont {T.}~\bibnamefont {Qi}}, \bibinfo {author} {\bibfnamefont
  {M.~A.}\ \bibnamefont {Hermele}}, \bibinfo {author} {\bibfnamefont
  {G.}~\bibnamefont {Cao}}, \ and\ \bibinfo {author} {\bibfnamefont {D.~S.}\
  \bibnamefont {Dessau}},\ }\href {\doibase 10.1038/ncomms11367} {\bibfield
  {journal} {\bibinfo  {journal} {Nat. Commun.}\ }\textbf {\bibinfo {volume}
  {7}},\ \bibinfo {pages} {11367} (\bibinfo {year} {2016})}\BibitemShut
  {NoStop}%
\bibitem [{\citenamefont {Qi}\ \emph {et~al.}(2012)\citenamefont {Qi},
  \citenamefont {Korneta}, \citenamefont {Li}, \citenamefont {Butrouna},
  \citenamefont {Cao}, \citenamefont {Wan}, \citenamefont {Schlottmann},
  \citenamefont {Kaul},\ and\ \citenamefont {Cao}}]{Qi2012}%
  \BibitemOpen
  \bibfield  {author} {\bibinfo {author} {\bibfnamefont {T.~F.}\ \bibnamefont
  {Qi}}, \bibinfo {author} {\bibfnamefont {O.~B.}\ \bibnamefont {Korneta}},
  \bibinfo {author} {\bibfnamefont {L.}~\bibnamefont {Li}}, \bibinfo {author}
  {\bibfnamefont {K.}~\bibnamefont {Butrouna}}, \bibinfo {author}
  {\bibfnamefont {V.~S.}\ \bibnamefont {Cao}}, \bibinfo {author} {\bibfnamefont
  {X.}~\bibnamefont {Wan}}, \bibinfo {author} {\bibfnamefont {P.}~\bibnamefont
  {Schlottmann}}, \bibinfo {author} {\bibfnamefont {R.~K.}\ \bibnamefont
  {Kaul}}, \ and\ \bibinfo {author} {\bibfnamefont {G.}~\bibnamefont {Cao}},\
  }\href {\doibase 10.1103/PhysRevB.86.125105} {\bibfield  {journal} {\bibinfo
  {journal} {Phys. Rev. B}\ }\textbf {\bibinfo {volume} {86}},\ \bibinfo
  {pages} {125105} (\bibinfo {year} {2012})}\BibitemShut {NoStop}%
\bibitem [{\citenamefont {Lee}\ \emph {et~al.}(2012)\citenamefont {Lee},
  \citenamefont {Krockenberger}, \citenamefont {Takahashi}, \citenamefont
  {Kawasaki},\ and\ \citenamefont {Tokura}}]{Lee2012}%
  \BibitemOpen
  \bibfield  {author} {\bibinfo {author} {\bibfnamefont {J.~S.}\ \bibnamefont
  {Lee}}, \bibinfo {author} {\bibfnamefont {Y.}~\bibnamefont {Krockenberger}},
  \bibinfo {author} {\bibfnamefont {K.~S.}\ \bibnamefont {Takahashi}}, \bibinfo
  {author} {\bibfnamefont {M.}~\bibnamefont {Kawasaki}}, \ and\ \bibinfo
  {author} {\bibfnamefont {Y.}~\bibnamefont {Tokura}},\ }\href {\doibase
  10.1103/PhysRevB.85.035101} {\bibfield  {journal} {\bibinfo  {journal} {Phys.
  Rev. B}\ }\textbf {\bibinfo {volume} {85}},\ \bibinfo {pages} {035101}
  (\bibinfo {year} {2012})}\BibitemShut {NoStop}%
\bibitem [{\citenamefont {Cetin}\ \emph {et~al.}(2012)\citenamefont {Cetin},
  \citenamefont {Lemmens}, \citenamefont {Gnezdilov}, \citenamefont
  {Wulferding}, \citenamefont {Menzel}, \citenamefont {Takayama}, \citenamefont
  {Ohashi},\ and\ \citenamefont {Takagi}}]{Cetin2012}%
  \BibitemOpen
  \bibfield  {author} {\bibinfo {author} {\bibfnamefont {M.~F.}\ \bibnamefont
  {Cetin}}, \bibinfo {author} {\bibfnamefont {P.}~\bibnamefont {Lemmens}},
  \bibinfo {author} {\bibfnamefont {V.}~\bibnamefont {Gnezdilov}}, \bibinfo
  {author} {\bibfnamefont {D.}~\bibnamefont {Wulferding}}, \bibinfo {author}
  {\bibfnamefont {D.}~\bibnamefont {Menzel}}, \bibinfo {author} {\bibfnamefont
  {T.}~\bibnamefont {Takayama}}, \bibinfo {author} {\bibfnamefont
  {K.}~\bibnamefont {Ohashi}}, \ and\ \bibinfo {author} {\bibfnamefont
  {H.}~\bibnamefont {Takagi}},\ }\href {\doibase 10.1103/PhysRevB.85.195148}
  {\bibfield  {journal} {\bibinfo  {journal} {Phys. Rev. B}\ }\textbf {\bibinfo
  {volume} {85}},\ \bibinfo {pages} {195148} (\bibinfo {year}
  {2012})}\BibitemShut {NoStop}%
\bibitem [{\citenamefont {Wang}\ \emph {et~al.}(2013)\citenamefont {Wang},
  \citenamefont {Cao}, \citenamefont {Waugh}, \citenamefont {Park},
  \citenamefont {Qi}, \citenamefont {Korneta}, \citenamefont {Cao},\ and\
  \citenamefont {Dessau}}]{Wang2013}%
  \BibitemOpen
  \bibfield  {author} {\bibinfo {author} {\bibfnamefont {Q.}~\bibnamefont
  {Wang}}, \bibinfo {author} {\bibfnamefont {Y.}~\bibnamefont {Cao}}, \bibinfo
  {author} {\bibfnamefont {J.~A.}\ \bibnamefont {Waugh}}, \bibinfo {author}
  {\bibfnamefont {S.~R.}\ \bibnamefont {Park}}, \bibinfo {author}
  {\bibfnamefont {T.~F.}\ \bibnamefont {Qi}}, \bibinfo {author} {\bibfnamefont
  {O.~B.}\ \bibnamefont {Korneta}}, \bibinfo {author} {\bibfnamefont
  {G.}~\bibnamefont {Cao}}, \ and\ \bibinfo {author} {\bibfnamefont {D.~S.}\
  \bibnamefont {Dessau}},\ }\href {\doibase 10.1103/PhysRevB.87.245109}
  {\bibfield  {journal} {\bibinfo  {journal} {Phys. Rev. B}\ }\textbf {\bibinfo
  {volume} {87}},\ \bibinfo {pages} {245109} (\bibinfo {year}
  {2013})}\BibitemShut {NoStop}%
\bibitem [{\citenamefont {Yan}\ \emph {et~al.}(2015)\citenamefont {Yan},
  \citenamefont {Ren}, \citenamefont {Xu}, \citenamefont {Xie}, \citenamefont
  {Tao}, \citenamefont {Choi}, \citenamefont {Lee}, \citenamefont {Choi},
  \citenamefont {Zhang},\ and\ \citenamefont {Feng}}]{Yan2015}%
  \BibitemOpen
  \bibfield  {author} {\bibinfo {author} {\bibfnamefont {Y.~J.}\ \bibnamefont
  {Yan}}, \bibinfo {author} {\bibfnamefont {M.~Q.}\ \bibnamefont {Ren}},
  \bibinfo {author} {\bibfnamefont {H.~C.}\ \bibnamefont {Xu}}, \bibinfo
  {author} {\bibfnamefont {B.~P.}\ \bibnamefont {Xie}}, \bibinfo {author}
  {\bibfnamefont {R.}~\bibnamefont {Tao}}, \bibinfo {author} {\bibfnamefont
  {H.~Y.}\ \bibnamefont {Choi}}, \bibinfo {author} {\bibfnamefont
  {N.}~\bibnamefont {Lee}}, \bibinfo {author} {\bibfnamefont {Y.~J.}\
  \bibnamefont {Choi}}, \bibinfo {author} {\bibfnamefont {T.}~\bibnamefont
  {Zhang}}, \ and\ \bibinfo {author} {\bibfnamefont {D.~L.}\ \bibnamefont
  {Feng}},\ }\href {\doibase 10.1103/PhysRevX.5.041018} {\bibfield  {journal}
  {\bibinfo  {journal} {Phys. Rev. X}\ }\textbf {\bibinfo {volume} {5}},\
  \bibinfo {pages} {041018} (\bibinfo {year} {2015})}\BibitemShut {NoStop}%
\bibitem [{\citenamefont {Seo}\ \emph {et~al.}(2017)\citenamefont {Seo},
  \citenamefont {Ahn}, \citenamefont {Song}, \citenamefont {Chen},
  \citenamefont {Wilson},\ and\ \citenamefont {Moon}}]{Seo2017}%
  \BibitemOpen
  \bibfield  {author} {\bibinfo {author} {\bibfnamefont {J.~H.}\ \bibnamefont
  {Seo}}, \bibinfo {author} {\bibfnamefont {G.~H.}\ \bibnamefont {Ahn}},
  \bibinfo {author} {\bibfnamefont {S.~J.}\ \bibnamefont {Song}}, \bibinfo
  {author} {\bibfnamefont {X.}~\bibnamefont {Chen}}, \bibinfo {author}
  {\bibfnamefont {S.~D.}\ \bibnamefont {Wilson}}, \ and\ \bibinfo {author}
  {\bibfnamefont {S.~J.}\ \bibnamefont {Moon}},\ }\href {\doibase
  10.1038/s41598-017-10725-z} {\bibfield  {journal} {\bibinfo  {journal} {Sci.
  Rep.}\ }\textbf {\bibinfo {volume} {7}},\ \bibinfo {pages} {10494} (\bibinfo
  {year} {2017})}\BibitemShut {NoStop}%
\bibitem [{\citenamefont {Cao}\ \emph {et~al.}(2018)\citenamefont {Cao},
  \citenamefont {Terzic}, \citenamefont {Zhao}, \citenamefont {Zheng},
  \citenamefont {De~Long},\ and\ \citenamefont {Riseborough}}]{Cao2018}%
  \BibitemOpen
  \bibfield  {author} {\bibinfo {author} {\bibfnamefont {G.}~\bibnamefont
  {Cao}}, \bibinfo {author} {\bibfnamefont {J.}~\bibnamefont {Terzic}},
  \bibinfo {author} {\bibfnamefont {H.~D.}\ \bibnamefont {Zhao}}, \bibinfo
  {author} {\bibfnamefont {H.}~\bibnamefont {Zheng}}, \bibinfo {author}
  {\bibfnamefont {L.~E.}\ \bibnamefont {De~Long}}, \ and\ \bibinfo {author}
  {\bibfnamefont {P.~S.}\ \bibnamefont {Riseborough}},\ }\href {\doibase
  10.1103/PhysRevLett.120.017201} {\bibfield  {journal} {\bibinfo  {journal}
  {Phys. Rev. Lett.}\ }\textbf {\bibinfo {volume} {120}},\ \bibinfo {pages}
  {017201} (\bibinfo {year} {2018})}\BibitemShut {NoStop}%
\bibitem [{\citenamefont {Zhou}\ \emph {et~al.}(2017)\citenamefont {Zhou},
  \citenamefont {Jiang}, \citenamefont {Chen},\ and\ \citenamefont
  {Wang}}]{Zhou2017}%
  \BibitemOpen
  \bibfield  {author} {\bibinfo {author} {\bibfnamefont {S.}~\bibnamefont
  {Zhou}}, \bibinfo {author} {\bibfnamefont {K.}~\bibnamefont {Jiang}},
  \bibinfo {author} {\bibfnamefont {H.}~\bibnamefont {Chen}}, \ and\ \bibinfo
  {author} {\bibfnamefont {Z.}~\bibnamefont {Wang}},\ }\href {\doibase
  10.1103/PhysRevX.7.041018} {\bibfield  {journal} {\bibinfo  {journal} {Phys.
  Rev. X}\ }\textbf {\bibinfo {volume} {7}},\ \bibinfo {pages} {041018}
  (\bibinfo {year} {2017})}\BibitemShut {NoStop}%
\bibitem [{\citenamefont {Wang}\ \emph {et~al.}(2018)\citenamefont {Wang},
  \citenamefont {Bachar}, \citenamefont {Teyssier}, \citenamefont {Luo},
  \citenamefont {Rischau}, \citenamefont {Scheerer}, \citenamefont {de~la
  Torre}, \citenamefont {Perry}, \citenamefont {Baumberger},\ and\
  \citenamefont {van~der Marel}}]{Wang2018}%
  \BibitemOpen
  \bibfield  {author} {\bibinfo {author} {\bibfnamefont {K.}~\bibnamefont
  {Wang}}, \bibinfo {author} {\bibfnamefont {N.}~\bibnamefont {Bachar}},
  \bibinfo {author} {\bibfnamefont {J.}~\bibnamefont {Teyssier}}, \bibinfo
  {author} {\bibfnamefont {W.}~\bibnamefont {Luo}}, \bibinfo {author}
  {\bibfnamefont {C.~W.}\ \bibnamefont {Rischau}}, \bibinfo {author}
  {\bibfnamefont {G.}~\bibnamefont {Scheerer}}, \bibinfo {author}
  {\bibfnamefont {A.}~\bibnamefont {de~la Torre}}, \bibinfo {author}
  {\bibfnamefont {R.~S.}\ \bibnamefont {Perry}}, \bibinfo {author}
  {\bibfnamefont {F.}~\bibnamefont {Baumberger}}, \ and\ \bibinfo {author}
  {\bibfnamefont {D.}~\bibnamefont {van~der Marel}},\ }\href {\doibase
  10.1103/PhysRevB.98.045107} {\bibfield  {journal} {\bibinfo  {journal} {Phys.
  Rev. B}\ }\textbf {\bibinfo {volume} {98}},\ \bibinfo {pages} {045107}
  (\bibinfo {year} {2018})}\BibitemShut {NoStop}%
\bibitem [{\citenamefont {Louat}\ \emph {et~al.}(2018)\citenamefont {Louat},
  \citenamefont {Bert}, \citenamefont {Serrier-Garcia}, \citenamefont
  {Bertran}, \citenamefont {Le~F\`evre}, \citenamefont {Rault},\ and\
  \citenamefont {Brouet}}]{Louat2018}%
  \BibitemOpen
  \bibfield  {author} {\bibinfo {author} {\bibfnamefont {A.}~\bibnamefont
  {Louat}}, \bibinfo {author} {\bibfnamefont {F.}~\bibnamefont {Bert}},
  \bibinfo {author} {\bibfnamefont {L.}~\bibnamefont {Serrier-Garcia}},
  \bibinfo {author} {\bibfnamefont {F.}~\bibnamefont {Bertran}}, \bibinfo
  {author} {\bibfnamefont {P.}~\bibnamefont {Le~F\`evre}}, \bibinfo {author}
  {\bibfnamefont {J.}~\bibnamefont {Rault}}, \ and\ \bibinfo {author}
  {\bibfnamefont {V.}~\bibnamefont {Brouet}},\ }\href {\doibase
  10.1103/PhysRevB.97.161109} {\bibfield  {journal} {\bibinfo  {journal} {Phys.
  Rev. B}\ }\textbf {\bibinfo {volume} {97}},\ \bibinfo {pages} {161109}
  (\bibinfo {year} {2018})}\BibitemShut {NoStop}%
\bibitem [{\citenamefont {de~la Torre}\ \emph {et~al.}(2015)\citenamefont
  {de~la Torre}, \citenamefont {McKeown~Walker}, \citenamefont {Bruno},
  \citenamefont {Ricc\'o}, \citenamefont {Wang}, \citenamefont
  {Gutierrez~Lezama}, \citenamefont {Scheerer}, \citenamefont {Giriat},
  \citenamefont {Jaccard}, \citenamefont {Berthod}, \citenamefont {Kim},
  \citenamefont {Hoesch}, \citenamefont {Hunter}, \citenamefont {Perry},
  \citenamefont {Tamai},\ and\ \citenamefont {Baumberger}}]{DelaTorre2015}%
  \BibitemOpen
  \bibfield  {author} {\bibinfo {author} {\bibfnamefont {A.}~\bibnamefont
  {de~la Torre}}, \bibinfo {author} {\bibfnamefont {S.}~\bibnamefont
  {McKeown~Walker}}, \bibinfo {author} {\bibfnamefont {F.~Y.}\ \bibnamefont
  {Bruno}}, \bibinfo {author} {\bibfnamefont {S.}~\bibnamefont {Ricc\'o}},
  \bibinfo {author} {\bibfnamefont {Z.}~\bibnamefont {Wang}}, \bibinfo {author}
  {\bibfnamefont {I.}~\bibnamefont {Gutierrez~Lezama}}, \bibinfo {author}
  {\bibfnamefont {G.}~\bibnamefont {Scheerer}}, \bibinfo {author}
  {\bibfnamefont {G.}~\bibnamefont {Giriat}}, \bibinfo {author} {\bibfnamefont
  {D.}~\bibnamefont {Jaccard}}, \bibinfo {author} {\bibfnamefont
  {C.}~\bibnamefont {Berthod}}, \bibinfo {author} {\bibfnamefont {T.~K.}\
  \bibnamefont {Kim}}, \bibinfo {author} {\bibfnamefont {M.}~\bibnamefont
  {Hoesch}}, \bibinfo {author} {\bibfnamefont {E.~C.}\ \bibnamefont {Hunter}},
  \bibinfo {author} {\bibfnamefont {R.~S.}\ \bibnamefont {Perry}}, \bibinfo
  {author} {\bibfnamefont {A.}~\bibnamefont {Tamai}}, \ and\ \bibinfo {author}
  {\bibfnamefont {F.}~\bibnamefont {Baumberger}},\ }\href {\doibase
  10.1103/PhysRevLett.115.176402} {\bibfield  {journal} {\bibinfo  {journal}
  {Phys. Rev. Lett.}\ }\textbf {\bibinfo {volume} {115}},\ \bibinfo {pages}
  {176402} (\bibinfo {year} {2015})}\BibitemShut {NoStop}%
\bibitem [{\citenamefont {Zwartsenberg}\ \emph {et~al.}()\citenamefont
  {Zwartsenberg}, \citenamefont {Day}, \citenamefont {Razzoli}, \citenamefont
  {Michiardi}, \citenamefont {Xu}, \citenamefont {Shi}, \citenamefont
  {Denlinger}, \citenamefont {Cao}, \citenamefont {Calder}, \citenamefont
  {Ueda}, \citenamefont {Bertinshaw}, \citenamefont {Takagi}, \citenamefont
  {Kim}, \citenamefont {Elfimov},\ and\ \citenamefont
  {Damascelli}}]{Zwartsenberg2019}%
  \BibitemOpen
  \bibfield  {author} {\bibinfo {author} {\bibfnamefont {B.}~\bibnamefont
  {Zwartsenberg}}, \bibinfo {author} {\bibfnamefont {R.~P.}\ \bibnamefont
  {Day}}, \bibinfo {author} {\bibfnamefont {E.}~\bibnamefont {Razzoli}},
  \bibinfo {author} {\bibfnamefont {M.}~\bibnamefont {Michiardi}}, \bibinfo
  {author} {\bibfnamefont {N.}~\bibnamefont {Xu}}, \bibinfo {author}
  {\bibfnamefont {M.}~\bibnamefont {Shi}}, \bibinfo {author} {\bibfnamefont
  {J.~D.}\ \bibnamefont {Denlinger}}, \bibinfo {author} {\bibfnamefont
  {G.}~\bibnamefont {Cao}}, \bibinfo {author} {\bibfnamefont {S.}~\bibnamefont
  {Calder}}, \bibinfo {author} {\bibfnamefont {K.}~\bibnamefont {Ueda}},
  \bibinfo {author} {\bibfnamefont {J.}~\bibnamefont {Bertinshaw}}, \bibinfo
  {author} {\bibfnamefont {H.}~\bibnamefont {Takagi}}, \bibinfo {author}
  {\bibfnamefont {B.}~\bibnamefont {Kim}}, \bibinfo {author} {\bibfnamefont
  {I.~S.}\ \bibnamefont {Elfimov}}, \ and\ \bibinfo {author} {\bibfnamefont
  {A.}~\bibnamefont {Damascelli}},\ }\href@noop {} {\enquote {\bibinfo {title}
  {{Spin-orbit-controlled metal-insulator transition in Sr$_2$IrO$_4$}},}\
  }\Eprint {http://arxiv.org/abs/arXiv:1903.00484 (2019)} {arXiv:1903.00484
  (2019)} \BibitemShut {NoStop}%
\bibitem [{\citenamefont {Ye}\ \emph {et~al.}(2013)\citenamefont {Ye},
  \citenamefont {Chi}, \citenamefont {Chakoumakos}, \citenamefont
  {Fernandez-Baca}, \citenamefont {Qi},\ and\ \citenamefont {Cao}}]{Ye2013}%
  \BibitemOpen
  \bibfield  {author} {\bibinfo {author} {\bibfnamefont {F.}~\bibnamefont
  {Ye}}, \bibinfo {author} {\bibfnamefont {S.}~\bibnamefont {Chi}}, \bibinfo
  {author} {\bibfnamefont {B.~C.}\ \bibnamefont {Chakoumakos}}, \bibinfo
  {author} {\bibfnamefont {J.~A.}\ \bibnamefont {Fernandez-Baca}}, \bibinfo
  {author} {\bibfnamefont {T.}~\bibnamefont {Qi}}, \ and\ \bibinfo {author}
  {\bibfnamefont {G.}~\bibnamefont {Cao}},\ }\href {\doibase
  10.1103/PhysRevB.87.140406} {\bibfield  {journal} {\bibinfo  {journal} {Phys.
  Rev. B}\ }\textbf {\bibinfo {volume} {87}},\ \bibinfo {pages} {140406}
  (\bibinfo {year} {2013})}\BibitemShut {NoStop}%
\bibitem [{\citenamefont {Ye}\ \emph {et~al.}(2015)\citenamefont {Ye},
  \citenamefont {Wang}, \citenamefont {Hoffmann}, \citenamefont {Wang},
  \citenamefont {Chi}, \citenamefont {Matsuda}, \citenamefont {Chakoumakos},
  \citenamefont {Fernandez-Baca},\ and\ \citenamefont {Cao}}]{Ye2015}%
  \BibitemOpen
  \bibfield  {author} {\bibinfo {author} {\bibfnamefont {F.}~\bibnamefont
  {Ye}}, \bibinfo {author} {\bibfnamefont {X.}~\bibnamefont {Wang}}, \bibinfo
  {author} {\bibfnamefont {C.}~\bibnamefont {Hoffmann}}, \bibinfo {author}
  {\bibfnamefont {J.}~\bibnamefont {Wang}}, \bibinfo {author} {\bibfnamefont
  {S.}~\bibnamefont {Chi}}, \bibinfo {author} {\bibfnamefont {M.}~\bibnamefont
  {Matsuda}}, \bibinfo {author} {\bibfnamefont {B.~C.}\ \bibnamefont
  {Chakoumakos}}, \bibinfo {author} {\bibfnamefont {J.~A.}\ \bibnamefont
  {Fernandez-Baca}}, \ and\ \bibinfo {author} {\bibfnamefont {G.}~\bibnamefont
  {Cao}},\ }\href {\doibase 10.1103/PhysRevB.92.201112} {\bibfield  {journal}
  {\bibinfo  {journal} {Phys. Rev. B}\ }\textbf {\bibinfo {volume} {92}},\
  \bibinfo {pages} {201112} (\bibinfo {year} {2015})}\BibitemShut {NoStop}%
\bibitem [{\citenamefont {Brouet}\ \emph {et~al.}(2015)\citenamefont {Brouet},
  \citenamefont {Mansart}, \citenamefont {Perfetti}, \citenamefont {Piovera},
  \citenamefont {Vobornik}, \citenamefont {Le~F\`evre}, \citenamefont
  {Bertran}, \citenamefont {Riggs}, \citenamefont {Shapiro}, \citenamefont
  {Giraldo-Gallo},\ and\ \citenamefont {Fisher}}]{Brouet2015}%
  \BibitemOpen
  \bibfield  {author} {\bibinfo {author} {\bibfnamefont {V.}~\bibnamefont
  {Brouet}}, \bibinfo {author} {\bibfnamefont {J.}~\bibnamefont {Mansart}},
  \bibinfo {author} {\bibfnamefont {L.}~\bibnamefont {Perfetti}}, \bibinfo
  {author} {\bibfnamefont {C.}~\bibnamefont {Piovera}}, \bibinfo {author}
  {\bibfnamefont {I.}~\bibnamefont {Vobornik}}, \bibinfo {author}
  {\bibfnamefont {P.}~\bibnamefont {Le~F\`evre}}, \bibinfo {author}
  {\bibfnamefont {F.~m.~c.}\ \bibnamefont {Bertran}}, \bibinfo {author}
  {\bibfnamefont {S.~C.}\ \bibnamefont {Riggs}}, \bibinfo {author}
  {\bibfnamefont {M.~C.}\ \bibnamefont {Shapiro}}, \bibinfo {author}
  {\bibfnamefont {P.}~\bibnamefont {Giraldo-Gallo}}, \ and\ \bibinfo {author}
  {\bibfnamefont {I.~R.}\ \bibnamefont {Fisher}},\ }\href {\doibase
  10.1103/PhysRevB.92.081117} {\bibfield  {journal} {\bibinfo  {journal} {Phys.
  Rev. B}\ }\textbf {\bibinfo {volume} {92}},\ \bibinfo {pages} {081117}
  (\bibinfo {year} {2015})}\BibitemShut {NoStop}%
\bibitem [{\citenamefont {Zhao}\ \emph {et~al.}(2016)\citenamefont {Zhao},
  \citenamefont {Torchinsky}, \citenamefont {Chu}, \citenamefont {Ivanov},
  \citenamefont {Lifshitz}, \citenamefont {Flint}, \citenamefont {Qi},
  \citenamefont {Cao},\ and\ \citenamefont {Hsieh}}]{Zhao2016}%
  \BibitemOpen
  \bibfield  {author} {\bibinfo {author} {\bibfnamefont {L.}~\bibnamefont
  {Zhao}}, \bibinfo {author} {\bibfnamefont {D.~H.}\ \bibnamefont
  {Torchinsky}}, \bibinfo {author} {\bibfnamefont {H.}~\bibnamefont {Chu}},
  \bibinfo {author} {\bibfnamefont {V.}~\bibnamefont {Ivanov}}, \bibinfo
  {author} {\bibfnamefont {R.}~\bibnamefont {Lifshitz}}, \bibinfo {author}
  {\bibfnamefont {R.}~\bibnamefont {Flint}}, \bibinfo {author} {\bibfnamefont
  {T.}~\bibnamefont {Qi}}, \bibinfo {author} {\bibfnamefont {G.}~\bibnamefont
  {Cao}}, \ and\ \bibinfo {author} {\bibfnamefont {D.}~\bibnamefont {Hsieh}},\
  }\href {\doibase 10.1038/nphys3517} {\bibfield  {journal} {\bibinfo
  {journal} {Nat. Phys.}\ }\textbf {\bibinfo {volume} {12}},\ \bibinfo {pages}
  {32} (\bibinfo {year} {2016})}\BibitemShut {NoStop}%
\bibitem [{\citenamefont {Jeong}\ \emph {et~al.}(2017)\citenamefont {Jeong},
  \citenamefont {Sidis}, \citenamefont {Louat}, \citenamefont {Brouet},\ and\
  \citenamefont {Bourges}}]{Jeong2017}%
  \BibitemOpen
  \bibfield  {author} {\bibinfo {author} {\bibfnamefont {J.}~\bibnamefont
  {Jeong}}, \bibinfo {author} {\bibfnamefont {Y.}~\bibnamefont {Sidis}},
  \bibinfo {author} {\bibfnamefont {A.}~\bibnamefont {Louat}}, \bibinfo
  {author} {\bibfnamefont {V.}~\bibnamefont {Brouet}}, \ and\ \bibinfo {author}
  {\bibfnamefont {P.}~\bibnamefont {Bourges}},\ }\href {\doibase
  10.1038/ncomms15119} {\bibfield  {journal} {\bibinfo  {journal} {Nat.
  Commun.}\ }\textbf {\bibinfo {volume} {8}},\ \bibinfo {pages} {15119}
  (\bibinfo {year} {2017})},\ \Eprint {http://arxiv.org/abs/1701.06485}
  {1701.06485} \BibitemShut {NoStop}%
\bibitem [{Note1()}]{Note1}%
  \BibitemOpen
  \bibinfo {note} {See Supplemental Material at \protect \url
  {http://link.aps.org/supplemental/xxx} for the details about the
  characterization of the samples, the reflectivity measurements and the
  Kramers-Kronig analysis, the Drude-Lorentz fits, as well as the details of
  temperature-dependent spectra, which includes Refs.~\cite
  {Homes1993,Dressel2002}.}\BibitemShut {Stop}%
\bibitem [{\citenamefont {Homes}\ \emph {et~al.}(1993)\citenamefont {Homes},
  \citenamefont {Reedyk}, \citenamefont {Cradles},\ and\ \citenamefont
  {Timusk}}]{Homes1993}%
  \BibitemOpen
  \bibfield  {author} {\bibinfo {author} {\bibfnamefont {C.~C.}\ \bibnamefont
  {Homes}}, \bibinfo {author} {\bibfnamefont {M.}~\bibnamefont {Reedyk}},
  \bibinfo {author} {\bibfnamefont {D.~A.}\ \bibnamefont {Cradles}}, \ and\
  \bibinfo {author} {\bibfnamefont {T.}~\bibnamefont {Timusk}},\ }\href
  {\doibase 10.1364/AO.32.002976} {\bibfield  {journal} {\bibinfo  {journal}
  {Appl. Opt.}\ }\textbf {\bibinfo {volume} {32}},\ \bibinfo {pages} {2976}
  (\bibinfo {year} {1993})}\BibitemShut {NoStop}%
\bibitem [{\citenamefont {Dressel}\ and\ \citenamefont
  {Gr\"uner}(2002)}]{Dressel2002}%
  \BibitemOpen
  \bibfield  {author} {\bibinfo {author} {\bibfnamefont {M.}~\bibnamefont
  {Dressel}}\ and\ \bibinfo {author} {\bibfnamefont {G.}~\bibnamefont
  {Gr\"uner}},\ }\href@noop {} {\emph {\bibinfo {title} {{Electrodynamics of
  Solids}}}}\ (\bibinfo  {publisher} {Cambridge University press},\ \bibinfo
  {year} {2002})\BibitemShut {NoStop}%
\bibitem [{\citenamefont {Moon}\ \emph {et~al.}(2009)\citenamefont {Moon},
  \citenamefont {Jin}, \citenamefont {Choi}, \citenamefont {Lee}, \citenamefont
  {Seo}, \citenamefont {Yu}, \citenamefont {Cao}, \citenamefont {Noh},\ and\
  \citenamefont {Lee}}]{Moon2009}%
  \BibitemOpen
  \bibfield  {author} {\bibinfo {author} {\bibfnamefont {S.~J.}\ \bibnamefont
  {Moon}}, \bibinfo {author} {\bibfnamefont {H.}~\bibnamefont {Jin}}, \bibinfo
  {author} {\bibfnamefont {W.~S.}\ \bibnamefont {Choi}}, \bibinfo {author}
  {\bibfnamefont {J.~S.}\ \bibnamefont {Lee}}, \bibinfo {author} {\bibfnamefont
  {S.~S.~A.}\ \bibnamefont {Seo}}, \bibinfo {author} {\bibfnamefont
  {J.}~\bibnamefont {Yu}}, \bibinfo {author} {\bibfnamefont {G.}~\bibnamefont
  {Cao}}, \bibinfo {author} {\bibfnamefont {T.~W.}\ \bibnamefont {Noh}}, \ and\
  \bibinfo {author} {\bibfnamefont {Y.~S.}\ \bibnamefont {Lee}},\ }\href
  {\doibase 10.1103/PhysRevB.80.195110} {\bibfield  {journal} {\bibinfo
  {journal} {Phys. Rev. B}\ }\textbf {\bibinfo {volume} {80}},\ \bibinfo
  {pages} {195110} (\bibinfo {year} {2009})}\BibitemShut {NoStop}%
\bibitem [{\citenamefont {Pr\"opper}\ \emph {et~al.}(2016)\citenamefont
  {Pr\"opper}, \citenamefont {Yaresko}, \citenamefont {H\"oppner},
  \citenamefont {Matiks}, \citenamefont {Mathis}, \citenamefont {Takayama},
  \citenamefont {Matsumoto}, \citenamefont {Takagi}, \citenamefont {Keimer},\
  and\ \citenamefont {Boris}}]{Propper2016}%
  \BibitemOpen
  \bibfield  {author} {\bibinfo {author} {\bibfnamefont {D.}~\bibnamefont
  {Pr\"opper}}, \bibinfo {author} {\bibfnamefont {A.~N.}\ \bibnamefont
  {Yaresko}}, \bibinfo {author} {\bibfnamefont {M.}~\bibnamefont {H\"oppner}},
  \bibinfo {author} {\bibfnamefont {Y.}~\bibnamefont {Matiks}}, \bibinfo
  {author} {\bibfnamefont {Y.-L.}\ \bibnamefont {Mathis}}, \bibinfo {author}
  {\bibfnamefont {T.}~\bibnamefont {Takayama}}, \bibinfo {author}
  {\bibfnamefont {A.}~\bibnamefont {Matsumoto}}, \bibinfo {author}
  {\bibfnamefont {H.}~\bibnamefont {Takagi}}, \bibinfo {author} {\bibfnamefont
  {B.}~\bibnamefont {Keimer}}, \ and\ \bibinfo {author} {\bibfnamefont {A.~V.}\
  \bibnamefont {Boris}},\ }\href {\doibase 10.1103/PhysRevB.94.035158}
  {\bibfield  {journal} {\bibinfo  {journal} {Phys. Rev. B}\ }\textbf {\bibinfo
  {volume} {94}},\ \bibinfo {pages} {035158} (\bibinfo {year}
  {2016})}\BibitemShut {NoStop}%
\bibitem [{\citenamefont {Souri}\ \emph {et~al.}(2017)\citenamefont {Souri},
  \citenamefont {Kim}, \citenamefont {Gruenewald}, \citenamefont {Connell},
  \citenamefont {Thompson}, \citenamefont {Nichols}, \citenamefont {Terzic},
  \citenamefont {Min}, \citenamefont {Cao}, \citenamefont {Brill},\ and\
  \citenamefont {Seo}}]{Souri2017}%
  \BibitemOpen
  \bibfield  {author} {\bibinfo {author} {\bibfnamefont {M.}~\bibnamefont
  {Souri}}, \bibinfo {author} {\bibfnamefont {B.~H.}\ \bibnamefont {Kim}},
  \bibinfo {author} {\bibfnamefont {J.~H.}\ \bibnamefont {Gruenewald}},
  \bibinfo {author} {\bibfnamefont {J.~G.}\ \bibnamefont {Connell}}, \bibinfo
  {author} {\bibfnamefont {J.}~\bibnamefont {Thompson}}, \bibinfo {author}
  {\bibfnamefont {J.}~\bibnamefont {Nichols}}, \bibinfo {author} {\bibfnamefont
  {J.}~\bibnamefont {Terzic}}, \bibinfo {author} {\bibfnamefont {B.~I.}\
  \bibnamefont {Min}}, \bibinfo {author} {\bibfnamefont {G.}~\bibnamefont
  {Cao}}, \bibinfo {author} {\bibfnamefont {J.~W.}\ \bibnamefont {Brill}}, \
  and\ \bibinfo {author} {\bibfnamefont {A.}~\bibnamefont {Seo}},\ }\href
  {\doibase 10.1103/PhysRevB.95.235125} {\bibfield  {journal} {\bibinfo
  {journal} {Phys. Rev. B}\ }\textbf {\bibinfo {volume} {95}},\ \bibinfo
  {pages} {235125} (\bibinfo {year} {2017})}\BibitemShut {NoStop}%
\bibitem [{\citenamefont {Falck}\ \emph {et~al.}(1992)\citenamefont {Falck},
  \citenamefont {Levy}, \citenamefont {Kastner},\ and\ \citenamefont
  {Birgeneau}}]{Falck1992}%
  \BibitemOpen
  \bibfield  {author} {\bibinfo {author} {\bibfnamefont {J.~P.}\ \bibnamefont
  {Falck}}, \bibinfo {author} {\bibfnamefont {A.}~\bibnamefont {Levy}},
  \bibinfo {author} {\bibfnamefont {M.~A.}\ \bibnamefont {Kastner}}, \ and\
  \bibinfo {author} {\bibfnamefont {R.~J.}\ \bibnamefont {Birgeneau}},\ }\href
  {\doibase 10.1103/PhysRevLett.69.1109} {\bibfield  {journal} {\bibinfo
  {journal} {Phys. Rev. Lett.}\ }\textbf {\bibinfo {volume} {69}},\ \bibinfo
  {pages} {1109} (\bibinfo {year} {1992})}\BibitemShut {NoStop}%
\bibitem [{\citenamefont {Lee}\ \emph {et~al.}(2005)\citenamefont {Lee},
  \citenamefont {Segawa}, \citenamefont {Li}, \citenamefont {Padilla},
  \citenamefont {Dumm}, \citenamefont {Dordevic}, \citenamefont {Homes},
  \citenamefont {Ando},\ and\ \citenamefont {Basov}}]{Lee2005}%
  \BibitemOpen
  \bibfield  {author} {\bibinfo {author} {\bibfnamefont {Y.~S.}\ \bibnamefont
  {Lee}}, \bibinfo {author} {\bibfnamefont {K.}~\bibnamefont {Segawa}},
  \bibinfo {author} {\bibfnamefont {Z.~Q.}\ \bibnamefont {Li}}, \bibinfo
  {author} {\bibfnamefont {W.~J.}\ \bibnamefont {Padilla}}, \bibinfo {author}
  {\bibfnamefont {M.}~\bibnamefont {Dumm}}, \bibinfo {author} {\bibfnamefont
  {S.~V.}\ \bibnamefont {Dordevic}}, \bibinfo {author} {\bibfnamefont {C.~C.}\
  \bibnamefont {Homes}}, \bibinfo {author} {\bibfnamefont {Y.}~\bibnamefont
  {Ando}}, \ and\ \bibinfo {author} {\bibfnamefont {D.~N.}\ \bibnamefont
  {Basov}},\ }\href {\doibase 10.1103/PhysRevB.72.054529} {\bibfield  {journal}
  {\bibinfo  {journal} {Phys. Rev. B}\ }\textbf {\bibinfo {volume} {72}},\
  \bibinfo {pages} {054529} (\bibinfo {year} {2005})}\BibitemShut {NoStop}%
\bibitem [{\citenamefont {Lupi}\ \emph {et~al.}(2009)\citenamefont {Lupi},
  \citenamefont {Nicoletti}, \citenamefont {Limaj}, \citenamefont
  {Baldassarre}, \citenamefont {Ortolani}, \citenamefont {Ono}, \citenamefont
  {Ando},\ and\ \citenamefont {Calvani}}]{Lupi2009}%
  \BibitemOpen
  \bibfield  {author} {\bibinfo {author} {\bibfnamefont {S.}~\bibnamefont
  {Lupi}}, \bibinfo {author} {\bibfnamefont {D.}~\bibnamefont {Nicoletti}},
  \bibinfo {author} {\bibfnamefont {O.}~\bibnamefont {Limaj}}, \bibinfo
  {author} {\bibfnamefont {L.}~\bibnamefont {Baldassarre}}, \bibinfo {author}
  {\bibfnamefont {M.}~\bibnamefont {Ortolani}}, \bibinfo {author}
  {\bibfnamefont {S.}~\bibnamefont {Ono}}, \bibinfo {author} {\bibfnamefont
  {Y.}~\bibnamefont {Ando}}, \ and\ \bibinfo {author} {\bibfnamefont
  {P.}~\bibnamefont {Calvani}},\ }\href {\doibase
  10.1103/PhysRevLett.102.206409} {\bibfield  {journal} {\bibinfo  {journal}
  {Phys. Rev. Lett.}\ }\textbf {\bibinfo {volume} {102}},\ \bibinfo {pages}
  {206409} (\bibinfo {year} {2009})}\BibitemShut {NoStop}%
\bibitem [{\citenamefont {Uchida}\ \emph {et~al.}(1991)\citenamefont {Uchida},
  \citenamefont {Ido}, \citenamefont {Takagi}, \citenamefont {Arima},
  \citenamefont {Tokura},\ and\ \citenamefont {Tajima}}]{Uchida1991}%
  \BibitemOpen
  \bibfield  {author} {\bibinfo {author} {\bibfnamefont {S.}~\bibnamefont
  {Uchida}}, \bibinfo {author} {\bibfnamefont {T.}~\bibnamefont {Ido}},
  \bibinfo {author} {\bibfnamefont {H.}~\bibnamefont {Takagi}}, \bibinfo
  {author} {\bibfnamefont {T.}~\bibnamefont {Arima}}, \bibinfo {author}
  {\bibfnamefont {Y.}~\bibnamefont {Tokura}}, \ and\ \bibinfo {author}
  {\bibfnamefont {S.}~\bibnamefont {Tajima}},\ }\href {\doibase
  10.1103/PhysRevB.43.7942} {\bibfield  {journal} {\bibinfo  {journal} {Phys.
  Rev. B}\ }\textbf {\bibinfo {volume} {43}},\ \bibinfo {pages} {7942}
  (\bibinfo {year} {1991})}\BibitemShut {NoStop}%
\bibitem [{\citenamefont {Stephan}\ and\ \citenamefont
  {Horsch}(1990)}]{Stephan1990}%
  \BibitemOpen
  \bibfield  {author} {\bibinfo {author} {\bibfnamefont {W.}~\bibnamefont
  {Stephan}}\ and\ \bibinfo {author} {\bibfnamefont {P.}~\bibnamefont
  {Horsch}},\ }\href {\doibase 10.1103/PhysRevB.42.8736} {\bibfield  {journal}
  {\bibinfo  {journal} {Phys. Rev. B}\ }\textbf {\bibinfo {volume} {42}},\
  \bibinfo {pages} {8736} (\bibinfo {year} {1990})}\BibitemShut {NoStop}%
\bibitem [{\citenamefont {Dagotto}\ \emph {et~al.}(1992)\citenamefont
  {Dagotto}, \citenamefont {Moreo}, \citenamefont {Ortolani}, \citenamefont
  {Riera},\ and\ \citenamefont {Scalapino}}]{Dagotto1992}%
  \BibitemOpen
  \bibfield  {author} {\bibinfo {author} {\bibfnamefont {E.}~\bibnamefont
  {Dagotto}}, \bibinfo {author} {\bibfnamefont {A.}~\bibnamefont {Moreo}},
  \bibinfo {author} {\bibfnamefont {F.}~\bibnamefont {Ortolani}}, \bibinfo
  {author} {\bibfnamefont {J.}~\bibnamefont {Riera}}, \ and\ \bibinfo {author}
  {\bibfnamefont {D.~J.}\ \bibnamefont {Scalapino}},\ }\href {\doibase
  10.1103/PhysRevB.45.10107} {\bibfield  {journal} {\bibinfo  {journal} {Phys.
  Rev. B}\ }\textbf {\bibinfo {volume} {45}},\ \bibinfo {pages} {10107}
  (\bibinfo {year} {1992})}\BibitemShut {NoStop}%
\bibitem [{\citenamefont {Dagotto}(1994)}]{Dagotto1994}%
  \BibitemOpen
  \bibfield  {author} {\bibinfo {author} {\bibfnamefont {E.}~\bibnamefont
  {Dagotto}},\ }\href {\doibase 10.1103/RevModPhys.66.763} {\bibfield
  {journal} {\bibinfo  {journal} {Rev. Mod. Phys.}\ }\textbf {\bibinfo {volume}
  {66}},\ \bibinfo {pages} {763} (\bibinfo {year} {1994})}\BibitemShut
  {NoStop}%
\bibitem [{\citenamefont {Nakano}\ \emph {et~al.}(2007)\citenamefont {Nakano},
  \citenamefont {Takahashi},\ and\ \citenamefont {Imada}}]{Nakano2007}%
  \BibitemOpen
  \bibfield  {author} {\bibinfo {author} {\bibfnamefont {H.}~\bibnamefont
  {Nakano}}, \bibinfo {author} {\bibfnamefont {Y.}~\bibnamefont {Takahashi}}, \
  and\ \bibinfo {author} {\bibfnamefont {M.}~\bibnamefont {Imada}},\ }\href
  {\doibase 10.1143/JPSJ.76.034705} {\bibfield  {journal} {\bibinfo  {journal}
  {Journal of the Physical Society of Japan}\ }\textbf {\bibinfo {volume}
  {76}},\ \bibinfo {pages} {034705} (\bibinfo {year} {2007})}\BibitemShut
  {NoStop}%
\bibitem [{\citenamefont {Coldea}\ \emph {et~al.}(2001)\citenamefont {Coldea},
  \citenamefont {Hayden}, \citenamefont {Aeppli}, \citenamefont {Perring},
  \citenamefont {Frost}, \citenamefont {Mason}, \citenamefont {Cheong},\ and\
  \citenamefont {Fisk}}]{Coldea2001}%
  \BibitemOpen
  \bibfield  {author} {\bibinfo {author} {\bibfnamefont {R.}~\bibnamefont
  {Coldea}}, \bibinfo {author} {\bibfnamefont {S.~M.}\ \bibnamefont {Hayden}},
  \bibinfo {author} {\bibfnamefont {G.}~\bibnamefont {Aeppli}}, \bibinfo
  {author} {\bibfnamefont {T.~G.}\ \bibnamefont {Perring}}, \bibinfo {author}
  {\bibfnamefont {C.~D.}\ \bibnamefont {Frost}}, \bibinfo {author}
  {\bibfnamefont {T.~E.}\ \bibnamefont {Mason}}, \bibinfo {author}
  {\bibfnamefont {S.-W.}\ \bibnamefont {Cheong}}, \ and\ \bibinfo {author}
  {\bibfnamefont {Z.}~\bibnamefont {Fisk}},\ }\href {\doibase
  10.1103/PhysRevLett.86.5377} {\bibfield  {journal} {\bibinfo  {journal}
  {Phys. Rev. Lett.}\ }\textbf {\bibinfo {volume} {86}},\ \bibinfo {pages}
  {5377} (\bibinfo {year} {2001})}\BibitemShut {NoStop}%
\bibitem [{\citenamefont {Kim}\ \emph {et~al.}(2012{\natexlab{b}})\citenamefont
  {Kim}, \citenamefont {Casa}, \citenamefont {Upton}, \citenamefont {Gog},
  \citenamefont {Kim}, \citenamefont {Mitchell}, \citenamefont {van
  Veenendaal}, \citenamefont {Daghofer}, \citenamefont {van~den Brink},
  \citenamefont {Khaliullin},\ and\ \citenamefont {Kim}}]{Kim2012}%
  \BibitemOpen
  \bibfield  {author} {\bibinfo {author} {\bibfnamefont {J.}~\bibnamefont
  {Kim}}, \bibinfo {author} {\bibfnamefont {D.}~\bibnamefont {Casa}}, \bibinfo
  {author} {\bibfnamefont {M.~H.}\ \bibnamefont {Upton}}, \bibinfo {author}
  {\bibfnamefont {T.}~\bibnamefont {Gog}}, \bibinfo {author} {\bibfnamefont
  {Y.-J.}\ \bibnamefont {Kim}}, \bibinfo {author} {\bibfnamefont {J.~F.}\
  \bibnamefont {Mitchell}}, \bibinfo {author} {\bibfnamefont {M.}~\bibnamefont
  {van Veenendaal}}, \bibinfo {author} {\bibfnamefont {M.}~\bibnamefont
  {Daghofer}}, \bibinfo {author} {\bibfnamefont {J.}~\bibnamefont {van~den
  Brink}}, \bibinfo {author} {\bibfnamefont {G.}~\bibnamefont {Khaliullin}}, \
  and\ \bibinfo {author} {\bibfnamefont {B.~J.}\ \bibnamefont {Kim}},\ }\href
  {\doibase 10.1103/PhysRevLett.108.177003} {\bibfield  {journal} {\bibinfo
  {journal} {Phys. Rev. Lett.}\ }\textbf {\bibinfo {volume} {108}},\ \bibinfo
  {pages} {177003} (\bibinfo {year} {2012}{\natexlab{b}})}\BibitemShut
  {NoStop}%
\bibitem [{\citenamefont {Cappelluti}\ \emph {et~al.}(2007)\citenamefont
  {Cappelluti}, \citenamefont {Ciuchi},\ and\ \citenamefont
  {Fratini}}]{Cappelluti2007}%
  \BibitemOpen
  \bibfield  {author} {\bibinfo {author} {\bibfnamefont {E.}~\bibnamefont
  {Cappelluti}}, \bibinfo {author} {\bibfnamefont {S.}~\bibnamefont {Ciuchi}},
  \ and\ \bibinfo {author} {\bibfnamefont {S.}~\bibnamefont {Fratini}},\ }\href
  {\doibase 10.1103/PhysRevB.76.125111} {\bibfield  {journal} {\bibinfo
  {journal} {Phys. Rev. B}\ }\textbf {\bibinfo {volume} {76}},\ \bibinfo
  {pages} {125111} (\bibinfo {year} {2007})}\BibitemShut {NoStop}%
\bibitem [{\citenamefont {Cappelluti}\ \emph {et~al.}(2009)\citenamefont
  {Cappelluti}, \citenamefont {Ciuchi},\ and\ \citenamefont
  {Fratini}}]{Cappelluti2009}%
  \BibitemOpen
  \bibfield  {author} {\bibinfo {author} {\bibfnamefont {E.}~\bibnamefont
  {Cappelluti}}, \bibinfo {author} {\bibfnamefont {S.}~\bibnamefont {Ciuchi}},
  \ and\ \bibinfo {author} {\bibfnamefont {S.}~\bibnamefont {Fratini}},\ }\href
  {\doibase 10.1103/PhysRevB.79.012502} {\bibfield  {journal} {\bibinfo
  {journal} {Phys. Rev. B}\ }\textbf {\bibinfo {volume} {79}},\ \bibinfo
  {pages} {012502} (\bibinfo {year} {2009})}\BibitemShut {NoStop}%
\bibitem [{\citenamefont {Mishchenko}\ \emph {et~al.}(2008)\citenamefont
  {Mishchenko}, \citenamefont {Nagaosa}, \citenamefont {Shen}, \citenamefont
  {De~Filippis}, \citenamefont {Cataudella}, \citenamefont {Devereaux},
  \citenamefont {Bernhard}, \citenamefont {Kim},\ and\ \citenamefont
  {Zaanen}}]{Mishchenko2008}%
  \BibitemOpen
  \bibfield  {author} {\bibinfo {author} {\bibfnamefont {A.~S.}\ \bibnamefont
  {Mishchenko}}, \bibinfo {author} {\bibfnamefont {N.}~\bibnamefont {Nagaosa}},
  \bibinfo {author} {\bibfnamefont {Z.-X.}\ \bibnamefont {Shen}}, \bibinfo
  {author} {\bibfnamefont {G.}~\bibnamefont {De~Filippis}}, \bibinfo {author}
  {\bibfnamefont {V.}~\bibnamefont {Cataudella}}, \bibinfo {author}
  {\bibfnamefont {T.~P.}\ \bibnamefont {Devereaux}}, \bibinfo {author}
  {\bibfnamefont {C.}~\bibnamefont {Bernhard}}, \bibinfo {author}
  {\bibfnamefont {K.~W.}\ \bibnamefont {Kim}}, \ and\ \bibinfo {author}
  {\bibfnamefont {J.}~\bibnamefont {Zaanen}},\ }\href {\doibase
  10.1103/PhysRevLett.100.166401} {\bibfield  {journal} {\bibinfo  {journal}
  {Phys. Rev. Lett.}\ }\textbf {\bibinfo {volume} {100}},\ \bibinfo {pages}
  {166401} (\bibinfo {year} {2008})}\BibitemShut {NoStop}%
\bibitem [{\citenamefont {Takenaka}\ \emph {et~al.}(2002)\citenamefont
  {Takenaka}, \citenamefont {Shiozaki}, \citenamefont {Okuyama}, \citenamefont
  {Nohara}, \citenamefont {Osuka}, \citenamefont {Takayanagi},\ and\
  \citenamefont {Sugai}}]{Takenaka2002}%
  \BibitemOpen
  \bibfield  {author} {\bibinfo {author} {\bibfnamefont {K.}~\bibnamefont
  {Takenaka}}, \bibinfo {author} {\bibfnamefont {R.}~\bibnamefont {Shiozaki}},
  \bibinfo {author} {\bibfnamefont {S.}~\bibnamefont {Okuyama}}, \bibinfo
  {author} {\bibfnamefont {J.}~\bibnamefont {Nohara}}, \bibinfo {author}
  {\bibfnamefont {A.}~\bibnamefont {Osuka}}, \bibinfo {author} {\bibfnamefont
  {Y.}~\bibnamefont {Takayanagi}}, \ and\ \bibinfo {author} {\bibfnamefont
  {S.}~\bibnamefont {Sugai}},\ }\href {\doibase 10.1103/PhysRevB.65.092405}
  {\bibfield  {journal} {\bibinfo  {journal} {Phys. Rev. B}\ }\textbf {\bibinfo
  {volume} {65}},\ \bibinfo {pages} {092405} (\bibinfo {year}
  {2002})}\BibitemShut {NoStop}%
\bibitem [{\citenamefont {Hu}\ \emph {et~al.}(2008)\citenamefont {Hu},
  \citenamefont {Dong}, \citenamefont {Li}, \citenamefont {Li}, \citenamefont
  {Zheng}, \citenamefont {Chen}, \citenamefont {Luo},\ and\ \citenamefont
  {Wang}}]{Hu2008}%
  \BibitemOpen
  \bibfield  {author} {\bibinfo {author} {\bibfnamefont {W.~Z.}\ \bibnamefont
  {Hu}}, \bibinfo {author} {\bibfnamefont {J.}~\bibnamefont {Dong}}, \bibinfo
  {author} {\bibfnamefont {G.}~\bibnamefont {Li}}, \bibinfo {author}
  {\bibfnamefont {Z.}~\bibnamefont {Li}}, \bibinfo {author} {\bibfnamefont
  {P.}~\bibnamefont {Zheng}}, \bibinfo {author} {\bibfnamefont {G.~F.}\
  \bibnamefont {Chen}}, \bibinfo {author} {\bibfnamefont {J.~L.}\ \bibnamefont
  {Luo}}, \ and\ \bibinfo {author} {\bibfnamefont {N.~L.}\ \bibnamefont
  {Wang}},\ }\href {\doibase 10.1103/PhysRevLett.101.257005} {\bibfield
  {journal} {\bibinfo  {journal} {Phys. Rev. Lett.}\ }\textbf {\bibinfo
  {volume} {101}},\ \bibinfo {pages} {257005} (\bibinfo {year}
  {2008})}\BibitemShut {NoStop}%
\bibitem [{\citenamefont {Mallett}\ \emph {et~al.}(2017)\citenamefont
  {Mallett}, \citenamefont {Wang}, \citenamefont {Marsik}, \citenamefont
  {Sheveleva}, \citenamefont {Yazdi-Rizi}, \citenamefont {Tallon},
  \citenamefont {Adelmann}, \citenamefont {Wolf},\ and\ \citenamefont
  {Bernhard}}]{Mallett2017PRB}%
  \BibitemOpen
  \bibfield  {author} {\bibinfo {author} {\bibfnamefont {B.~P.~P.}\
  \bibnamefont {Mallett}}, \bibinfo {author} {\bibfnamefont {C.~N.}\
  \bibnamefont {Wang}}, \bibinfo {author} {\bibfnamefont {P.}~\bibnamefont
  {Marsik}}, \bibinfo {author} {\bibfnamefont {E.}~\bibnamefont {Sheveleva}},
  \bibinfo {author} {\bibfnamefont {M.}~\bibnamefont {Yazdi-Rizi}}, \bibinfo
  {author} {\bibfnamefont {J.~L.}\ \bibnamefont {Tallon}}, \bibinfo {author}
  {\bibfnamefont {P.}~\bibnamefont {Adelmann}}, \bibinfo {author}
  {\bibfnamefont {T.}~\bibnamefont {Wolf}}, \ and\ \bibinfo {author}
  {\bibfnamefont {C.}~\bibnamefont {Bernhard}},\ }\href {\doibase
  10.1103/PhysRevB.95.054512} {\bibfield  {journal} {\bibinfo  {journal} {Phys.
  Rev. B}\ }\textbf {\bibinfo {volume} {95}},\ \bibinfo {pages} {054512}
  (\bibinfo {year} {2017})}\BibitemShut {NoStop}%
\bibitem [{\citenamefont {Xu}\ \emph {et~al.}(2018)\citenamefont {Xu},
  \citenamefont {Xiao}, \citenamefont {Gao}, \citenamefont {Ma}, \citenamefont
  {Mu}, \citenamefont {Marsik}, \citenamefont {Sheveleva}, \citenamefont
  {Lyzwa}, \citenamefont {Dai}, \citenamefont {Lobo},\ and\ \citenamefont
  {Bernhard}}]{Xu2018}%
  \BibitemOpen
  \bibfield  {author} {\bibinfo {author} {\bibfnamefont {B.}~\bibnamefont
  {Xu}}, \bibinfo {author} {\bibfnamefont {H.}~\bibnamefont {Xiao}}, \bibinfo
  {author} {\bibfnamefont {B.}~\bibnamefont {Gao}}, \bibinfo {author}
  {\bibfnamefont {Y.~H.}\ \bibnamefont {Ma}}, \bibinfo {author} {\bibfnamefont
  {G.}~\bibnamefont {Mu}}, \bibinfo {author} {\bibfnamefont {P.}~\bibnamefont
  {Marsik}}, \bibinfo {author} {\bibfnamefont {E.}~\bibnamefont {Sheveleva}},
  \bibinfo {author} {\bibfnamefont {F.}~\bibnamefont {Lyzwa}}, \bibinfo
  {author} {\bibfnamefont {Y.~M.}\ \bibnamefont {Dai}}, \bibinfo {author}
  {\bibfnamefont {R.~P. S.~M.}\ \bibnamefont {Lobo}}, \ and\ \bibinfo {author}
  {\bibfnamefont {C.}~\bibnamefont {Bernhard}},\ }\href {\doibase
  10.1103/PhysRevB.97.195110} {\bibfield  {journal} {\bibinfo  {journal} {Phys.
  Rev. B}\ }\textbf {\bibinfo {volume} {97}},\ \bibinfo {pages} {195110}
  (\bibinfo {year} {2018})}\BibitemShut {NoStop}%
\bibitem [{\citenamefont {Charnukha}\ \emph {et~al.}(2013)\citenamefont
  {Charnukha}, \citenamefont {Pr\"opper}, \citenamefont {Larkin}, \citenamefont
  {Sun}, \citenamefont {Li}, \citenamefont {Lin}, \citenamefont {Wolf},
  \citenamefont {Keimer},\ and\ \citenamefont {Boris}}]{Charnukha2013}%
  \BibitemOpen
  \bibfield  {author} {\bibinfo {author} {\bibfnamefont {A.}~\bibnamefont
  {Charnukha}}, \bibinfo {author} {\bibfnamefont {D.}~\bibnamefont
  {Pr\"opper}}, \bibinfo {author} {\bibfnamefont {T.~I.}\ \bibnamefont
  {Larkin}}, \bibinfo {author} {\bibfnamefont {D.~L.}\ \bibnamefont {Sun}},
  \bibinfo {author} {\bibfnamefont {Z.~W.}\ \bibnamefont {Li}}, \bibinfo
  {author} {\bibfnamefont {C.~T.}\ \bibnamefont {Lin}}, \bibinfo {author}
  {\bibfnamefont {T.}~\bibnamefont {Wolf}}, \bibinfo {author} {\bibfnamefont
  {B.}~\bibnamefont {Keimer}}, \ and\ \bibinfo {author} {\bibfnamefont {A.~V.}\
  \bibnamefont {Boris}},\ }\href {\doibase 10.1103/PhysRevB.88.184511}
  {\bibfield  {journal} {\bibinfo  {journal} {Phys. Rev. B}\ }\textbf {\bibinfo
  {volume} {88}},\ \bibinfo {pages} {184511} (\bibinfo {year}
  {2013})}\BibitemShut {NoStop}%
\bibitem [{\citenamefont {Toschi}\ \emph {et~al.}(2005)\citenamefont {Toschi},
  \citenamefont {Capone}, \citenamefont {Ortolani}, \citenamefont {Calvani},
  \citenamefont {Lupi},\ and\ \citenamefont {Castellani}}]{Toschi2005}%
  \BibitemOpen
  \bibfield  {author} {\bibinfo {author} {\bibfnamefont {A.}~\bibnamefont
  {Toschi}}, \bibinfo {author} {\bibfnamefont {M.}~\bibnamefont {Capone}},
  \bibinfo {author} {\bibfnamefont {M.}~\bibnamefont {Ortolani}}, \bibinfo
  {author} {\bibfnamefont {P.}~\bibnamefont {Calvani}}, \bibinfo {author}
  {\bibfnamefont {S.}~\bibnamefont {Lupi}}, \ and\ \bibinfo {author}
  {\bibfnamefont {C.}~\bibnamefont {Castellani}},\ }\href {\doibase
  10.1103/PhysRevLett.95.097002} {\bibfield  {journal} {\bibinfo  {journal}
  {Phys. Rev. Lett.}\ }\textbf {\bibinfo {volume} {95}},\ \bibinfo {pages}
  {097002} (\bibinfo {year} {2005})}\BibitemShut {NoStop}%
\end{thebibliography}
\end{document}